\documentclass[twocolumn]{aastex61}

\usepackage{amsmath}

\received{---}
\revised{---}
\accepted{---}
\submitjournal{PASP}

\shorttitle{POLICAN}
\shortauthors{Devaraj et al.}
\begin{document}

\title{POLICAN: A near-infrared imaging polarimeter at the $2.1\,\mathrm{m}$ OAGH telescope}

\correspondingauthor{Devaraj Rangaswamy}
\email{dev2future@yahoo.com}

\author{Devaraj R.}
\affiliation{Instituto Nacional de Astrof\'isica, \'Optica y Electr\'onica, \\ Luis Enrique Erro \# 1, Tonantzintla, Puebla - 72840, M\'exico}

\author{A. Luna}
\affiliation{Instituto Nacional de Astrof\'isica, \'Optica y Electr\'onica, \\ Luis Enrique Erro \# 1, Tonantzintla, Puebla - 72840, M\'exico}

\author{L. Carrasco}
\affiliation{Instituto Nacional de Astrof\'isica, \'Optica y Electr\'onica, \\ Luis Enrique Erro \# 1, Tonantzintla, Puebla - 72840, M\'exico}

\author{M. A. V\'azquez-Rodr\'iguez}
\affiliation{Instituto Nacional de Astrof\'isica, \'Optica y Electr\'onica, \\ Luis Enrique Erro \# 1, Tonantzintla, Puebla - 72840, M\'exico}

\author{Y. D. Mayya}
\affiliation{Instituto Nacional de Astrof\'isica, \'Optica y Electr\'onica, \\ Luis Enrique Erro \# 1, Tonantzintla, Puebla - 72840, M\'exico}

\author{J. G. T\'anori}
\affiliation{Instituto Nacional de Astrof\'isica, \'Optica y Electr\'onica, \\ Luis Enrique Erro \# 1, Tonantzintla, Puebla - 72840, M\'exico}

\author{E. O. Serrano Bernal}
\affiliation{Instituto Nacional de Astrof\'isica, \'Optica y Electr\'onica, \\ Luis Enrique Erro \# 1, Tonantzintla, Puebla - 72840, M\'exico}

\begin{abstract}

POLICAN is a near-infrared imaging linear polarimeter developed for the Cananea Near-infrared Camera (CANICA) at the $2.1\,\mathrm{m}$ telescope of the Guillermo Haro Astrophysical Observatory (OAGH) located in Cananea, Sonora, M\'exico. POLICAN is mounted ahead of CANICA and consist of a rotating super-achromatic ($1-2.7\,\mathrm{\mu m}$) half-wave plate (HWP) as the modulator and a fixed wire-grid polarizer as the analyzer. CANICA has a 1024 $\times$ 1024 HgCdTe detector with a plate scale of $0.32\,\mathrm{arcsec/pixel}$ and provides a field of view of $5.5 \times 5.5\,\mathrm{arcmin^2}$. The polarimetric observations are carried out by modulating the incoming light through different steps of half-wave plate angles $(0^{\circ}, 22$.$5^{\circ}, 45^{\circ}, 67$.$5^{\circ})$, to establish linear Stokes parameters ($I, Q$, and $U$). Image reduction consists of dark subtraction, polarimetric flat fielding, and sky subtraction. The astrometry and photometric calibrations are performed using the publicly available data from the Two Micron All Sky Survey. Polarimetric calibration includes observations of globular clusters and polarization standards available in the literature. Analysis of multiple observations of globular clusters yielded an instrumental polarization of 0.51\%. Uncertainties in polarization range from 0.1\% to 10\% from the brightest $7\,\mathrm{mag}$ to faintest $16\,\mathrm{mag}$ stars. The polarimetric accuracy achieved is better than 0.5\% and the position angle errors less than $5^{\circ}$ for stars brighter than $13\,\mathrm{mag}$ in $H$-band. POLICAN is mainly being used to study the scattered polarization and magnetic fields in and around star-forming regions of the interstellar medium.

\end{abstract}



\section{Introduction}

Imaging polarimeters offer great opportunity to study various astrophysical topics ranging from galactic regions to extragalactic sources such as active galactic nuclei (AGNs). Polarimeters built to function in optical wavelength are plenty \citep[e.g. IAGPOL, IMPOL, Dipol-2, RoboPol;][]{mag96, ram98, pir14, king14}, but they are limited to provide only partial insight into some of the science cases. On the other hand, near-infrared (NIR) polarimetry offers a unique window to observe new regions, revealing different physical phenomena. One of the main subjects of interest for polarimetric studies with $2\,\mathrm{m}$ class telescopes is the interstellar dust and cool galactic star-forming regions. The linearly polarized light from the stars, caused by dichroic extinction \citep{hall49, hiltner49} from dust grains, which are aligned to local magnetic fields \citep{davis51,lazarain07}, is very useful in understanding the interplay of interstellar matter and magnetic fields. Combining theory and observations, we can begin to understand dust properties and magnetic fields from dense cores to star-forming regions in and around molecular cloud complexes \citep[e.g.][]{jones89, nishi09, chapman11}. Existing NIR polarimeters like SIRPOL \citep{kandori06} and Mimir \citep{clemens07} have been used to conduct various observations and surveys like GPIPS \citep{clemens12a} to study the magnetic fields in the galactic medium. Additionally, NIR polarimetry can aid in investigating the circumstellar structures of young stellar objects (YSOs) whose radiation is scattered by dust and is observed as infrared reflection nebulae \citep{tamura06, hashimoto08}. Polarization data from spiral galaxies can be used to study galaxian magnetic field properties and the field's orientation to the disk \citep[e.g.][]{jones97, clemens13, mont14}.

With all of these diverse astrophysical topics to exploit, and considering that only few NIR polarimeters are available, we built a new instrument called POLICAN\footnote{POLICAN: \underline{Pol}arimetro \underline{I}nfrarojo para \underline{CAN}ICA} to function as an imaging linear polarimeter. POLICAN operates at the $2.1\,\mathrm{m}$ telescope of the Guillermo Haro Astrophysical Observatory (OAGH) in Cananea, Sonora, M\'exico. It is attached to the Cananea Near-infrared Camera (CANICA) \citep{car17} to operate as one of the primary backend instruments at the telescope. The instrument development was completed in the year 2012 with the support of funding from the Mexican science agency CONACyT.

\begin{figure*}
\epsscale{1.1}
\plotone{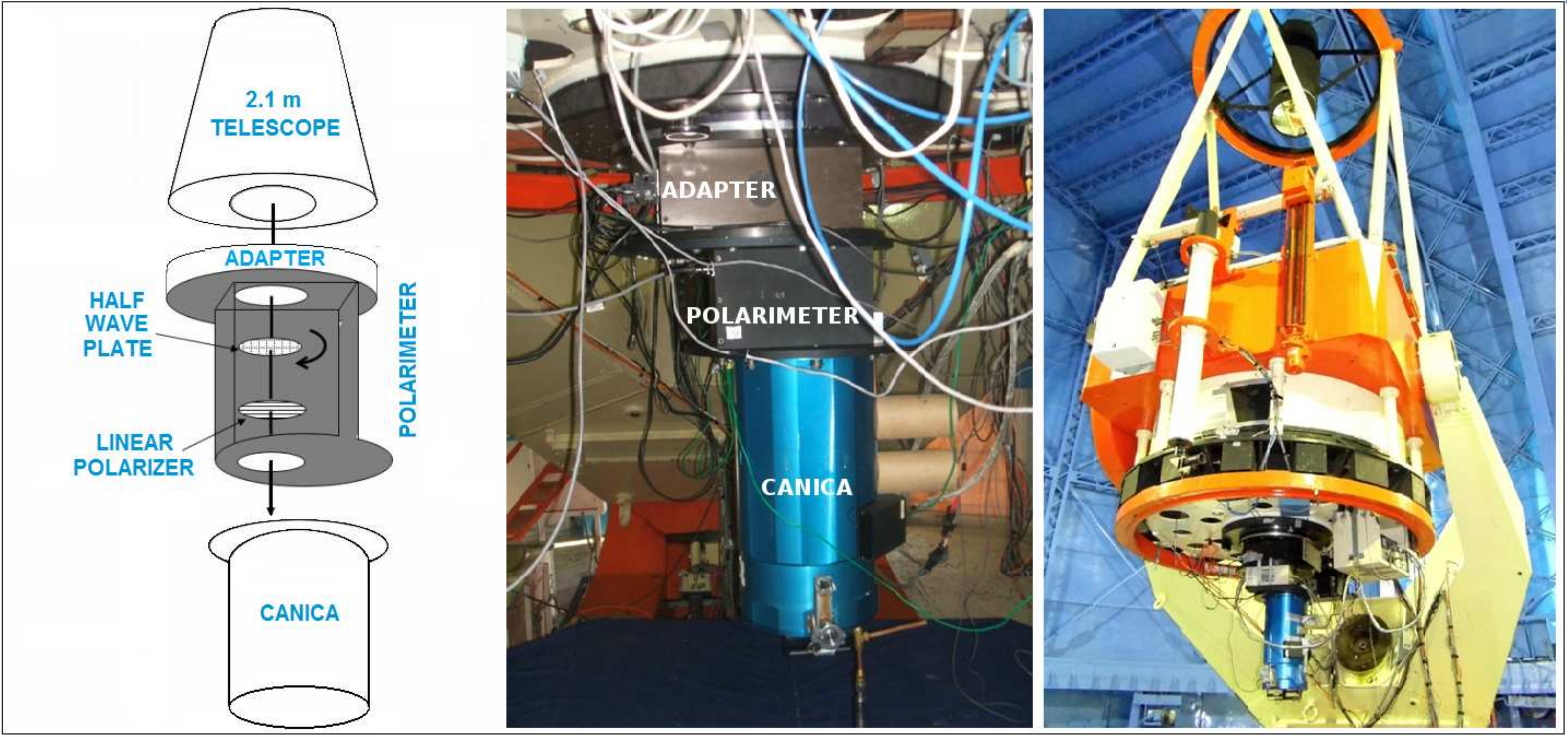} 
\caption{Block diagram and exterior photos of POLICAN attached to CANICA and the telescope. The polarimeter unit is a mechanical assembly consisting of a half-wave plate and a polarizer operated at room temperature. The assembly is coupled to CANICA on one end and to the telescope on the other end with the help of a rotating adapter.}
\label{fig1}
\end{figure*}

POLICAN consists of basic polarizing elements: a rotating super-achromatic ($1-2.7\,\mathrm{\mu m}$) half-wave plate (HWP) as the modulator and a fixed wire-grid polarizer as the analyzer. These are housed in an external assembly placed between the telescope and CANICA as shown in Figure~\ref{fig1}.

To meet the scientific requirements and to obtain good quality polarization data, it is important to make POLICAN function well and to understand its characteristics and operational behavior. Obtaining accurate polarimetric data requires optimization of observation methods and a robust data processing and analysis tool kit. Further, a detailed calibration of the instrument is necessary to calculate the true polarization. This led to development of various strategies and methods for operation and calibration of POLICAN. Additionally, software pipelines were developed for handling the large amounts of data to be processed into science-quality results. The core of POLICAN capabilities depends on the CANICA characteristics, which have been characterized and evaluated in \citet{dev17}; hereafter Paper~I. POLICAN incorporates a mechanical design similar to SIRPOL and most of its calibration scheme are derived from the Mimir team's approach \citep{clemens12b}. Preliminary descriptions about POLICAN are reported in \citet{dev15,dev17a}. In the following sections, we present the instrument overview, polarimeter operation, observational properties, data processing methods, calibration, and observational results of POLICAN.

\section{Instrument overview}

The reflector telescope at the OAGH observatory is a Ritchey-Chr\'etien configuration on an equatorial mount with a primary mirror of $2.1\,\mathrm{m}$ diameter and a secondary hyperbolic mirror of $50\,\mathrm{cm}$, yielding a focal length of $25\,\mathrm{m}$. The CANICA optics re-image the f/12 beam of the telescope to f/6 onto the detector. This optical setup provides observed images with an average point spread function (PSF) of $1.{\arcsec}5$ FWHM. CANICA operates primarily in the three broadbands $J (1.24\,\mathrm{\mu m})$, $H (1.63\,\mathrm{\mu m})$ and $K^{'} (2.12\,\mathrm{\mu m})$ including other multiple narrow bands. The camera unit is made up of a cryostat assembly including collimator, filter wheels, focusing system and a detector. The detector is a HgCdTe HAWAII array of 1024 $\times$ 1024 pixels with a plate scale of $0.32\,\mathrm{arcsec/pixel}$ and provides a field of view (FOV) of $5.5 \times 5.5\,\mathrm{arcmin^2}$. CANICA is operated using a correlated double sampling (CDS) readout method, and hence the raw data delivered after POLICAN observations are the CDS images. Details of CANICA construction and design are presented in \citet{car17}. A complete description of CANICA characteristics and performance are presented in Paper~I. 

\subsection{Polarimeter}

The CANICA has a light entrance window that is off-centered with respect to the cryostat so as to align it with the optical axis designed to accommodate filter wheels. To adjust for the displaced entrance window, an adapter couples the telescope and the CANICA. As a result, it was observed that the polarimeter POLICAN can be implemented in the space next to the adapter putting the polarizing elements externally to the CANICA at room temperature (see Figure~\ref{fig1}).

POLICAN's polarizing elements consists of a retarder (half-wave plate) and an analyzer (wire-grid polarizer). The retarder is a super-achromatic ($1-2.7\,\mathrm{\mu m}$) HWP of diameter $50\,\mathrm{mm}$, which is made by cementing pairs of MgF$_{2}$ and Quartz plates. It has very low path difference of $\pm 0.04\%$ and the change in orientation of the optical axis is negligible $\pm 0.2^{\circ}$ across the entire spectral range. The retarder is manufactured by Bernhard Halle Nachfl, Germany. The analyzer is a holographic high extinction ratio (HER) polarizer of diameter $71\,\mathrm{mm}$ deposited on a CaF$_{2}$ substrate, with a large spectral range from $1-10\,\mathrm{\mu m}$. It has a grid spacing of 4000~lines/mm with a transmission efficiency of 84\% at $2.5\,\mathrm{\mu m}$. The polarizer is manufactured by Specac company, UK.

\begin{figure*}
\epsscale{0.9}
\plotone{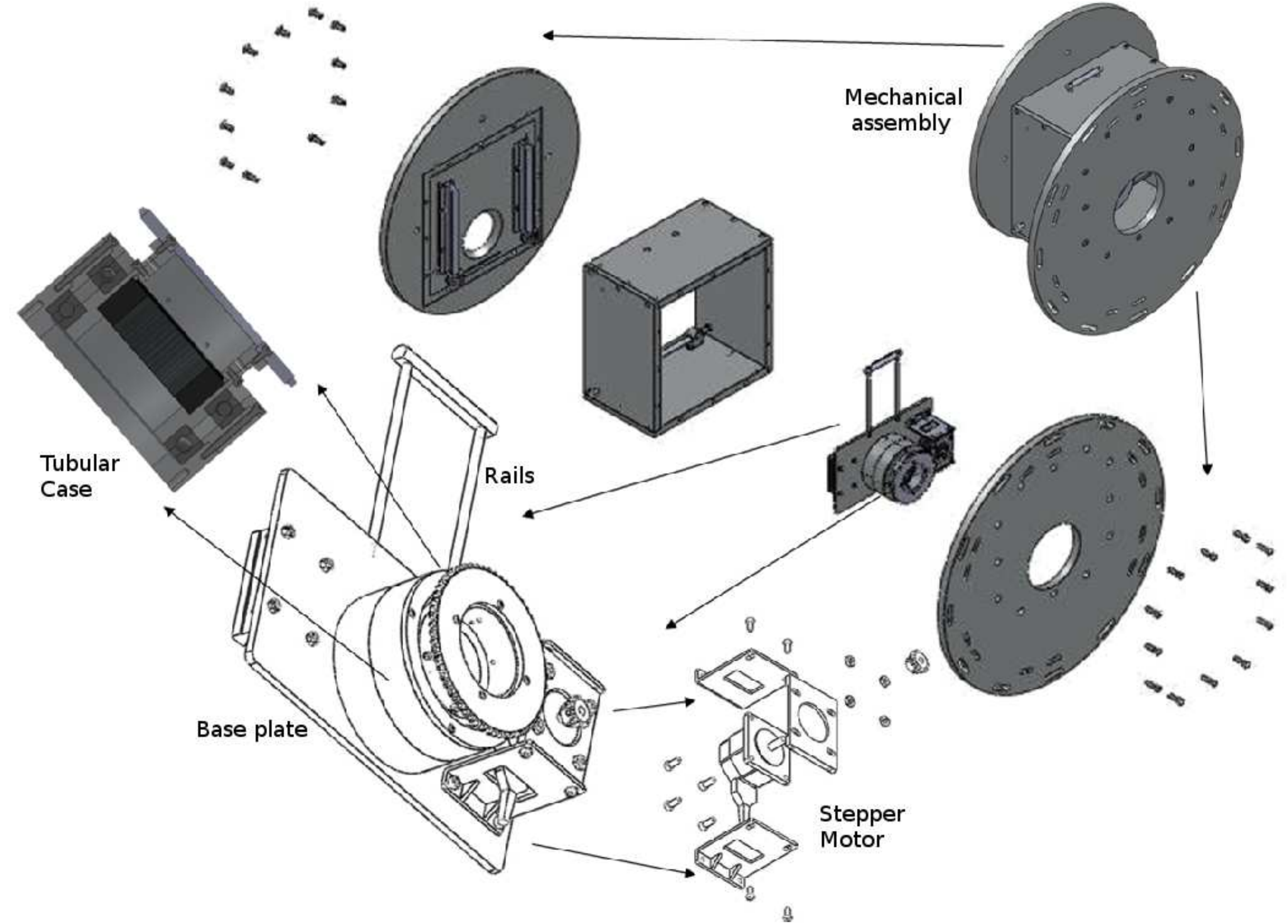} 
\caption{Blown-up view of the mechanical assembly of the polarimeter. The polarizing elements: half-wave plate and the polarizer are attached to a tubular case fixed onto a base plate. The half-wave plate is secured on a rotating unit connected to a stepper motor. The base plate is attached to set of rails that can move the unit in and out of the light path manually.}
\label{fig2}
\end{figure*}

The polarizing elements are housed inside an aluminum mechanical assembly with a detachable system. The detachable system can be manually removed from the beam along a translating stage, so as to switch the observations between normal photometry and imaging polarimetry. The change in back focal length during each mode is corrected by positioning the secondary mirror accordingly. The top and bottom ends of the mechanical assembly are provisioned with circular flanges for attaching to the telescope and to CANICA. The flexure in the mechanical assembly due to the weight of CANICA at different declinations is found to be negligible. The adapter for coupling POLICAN with the telescope is a rotating system that can be used to orient the instrument. The entire setup with the adapter is aligned to a setting of 328$^{\circ}$ to orient the observations along north-up and east-right direction on the detector. Figure~\ref{fig2} shows a zoomed view of the mechanical assembly with the stepper motor, HWP, and the polarizer. Details of the mechanical assembly design and construction are presented in \citet{vazrod12}.

\section{Operation and Control}\label{opcntrl}

Linear polarimetric observations can be achieved from a combination of rotating modulators and analyzers whereby the orthogonal components of polarization are produced. Dual-beam polarimeters \citep[e.g. PLANETPOL, DBIP, MMTPOL;][]{hough06, masiero07, pack12} that have a Glan-Thompson prism or a Wollaston prism as analyzers can produce two orthogonal polarizing components simultaneously for a single position of the modulator, thereby operating faster and reducing sky-dependent noise. However, they are restricted to observing well separated or isolated sources across a fairly narrower field. POLICAN, on the other hand, is designed to operate in a single-beam mode which has the advantage of observing both point and extended sources with a medium FOV. However, it must compromise on time cadence and deal with the effects of sky-dependent noise. Such a design of POLICAN requires a minimum of four modulation angles to obtain the polarizing components needed to establish the linear Stokes parameters \citep{shur62}. The change in the polarization state of the light as it passes through different optical components can be described by a Mueller matrix formalism \citep{clarke10}. From analysis of Mueller matrices, we find that the input linear Stokes parameters $I$, $Q$, and $U$ can be established from the output lights intensity, by modulating the incoming light with four HWP angles of the first quadrant$\mbox{---}0^{\circ}, 22$.$5^{\circ}, 45^{\circ}, 67$.$5^{\circ}$.  Appendix~\ref{mlrmat} describes the detail derivation of the modulation scheme with the use of Mueller matrices. Other sets of HWP angles from different quadrants can be used similarly to obtain the Stokes parameters. The results from each quadrant can then be combined to reduce HWP dependent errors. However, with POLICAN setup, we chose only the angles of the first quadrant to limit the large integration time during observations.
 
The control for HWP rotation for modulation is handled by a stepper motor that is integrated into the polarimeter mechanical assembly. The stepper motor connects to the HWP with a gear-to-gear transmission system and the rotation delivered is followed in precise step angle of 0.3$^{\circ}$. The reference home position of the HWP (i.e.\ the zero-phase axis) is identified by a Hall effect sensor. Connection to the motor controller from the main observation computer is through a RS232 serial communication line. The stepper motor unit is provided by Parker Motion Control Systems, USA, and the motor controller program is a software in the Visual BASIC. To simultaneously achieve control of CANICA and POLICAN, the stepper motor control and the CANICA control are integrated. This allows for scripted observations to be acquired in sequence for each HWP angle at each dithered position. 

\section{Observation goals}\label{obgoals}
The chief scientific goal of POLICAN is to study magnetic fields in the star-forming regions of the nearby (distances $\sim$few kpc) interstellar medium (ISM). Hence, it was important to define various observational properties to carry out polarimetric studies meeting the scientific requirements.

\subsection{Area coverage}
Most of the polarimeters available in the NIR can either do wide-field, large-scale surveys (e.g.\ Mimir, SIRPOL), spanning a few tens of parsecs, or use adaptive optics to study narrow field regions \citep[e.g. ZIMPOL, GPI;][]{rol10, perrin15}, across scales of $100-1000\,\mathrm{AU}$. With POLICAN we aim to bridge this gap to obtain polarization information at intermediate scales, between few-parsec to sub-parsec scales in the ISM. On the spatial range, we can achieve this by targeting areas of size from $\sim0.2\,\mathrm{arcmin^2}$ to $200\,\mathrm{arcmin^2}$. 

\subsection{Waveband}
At NIR bands, the magnetic field information in the ISM obtained from starlight polarization is revealed due to dichroic extinction. While $K$-band offers the best window to probe regions of high extinction ($A_{V}$), the thermal emission from the sky is large. On the other hand, at $J$-band the sky emission is low, but the $A_{V}$ values probed are limited. The $H$-band offers the best compromise, where we can sufficiently probe regions of moderate $A_{V}$ with low thermal emission from the sky. Hence, current observations with POLICAN are concentrated in $H$-band, which are presented in this article. However, calibration and study of sources in $J$ and $K$-band is being carried out simultaneously. 

\subsection{Sensitivity and sampling goals}
The starlight polarization information from the reddened stars in the ISM is usually weak, of the order of $1\%-5\%$ \citep{mat70}. To accurately perform starlight polarimetry for studying magnetic field properties requires measurements with polarization uncertainties below 1\% \citep{clemens12a}. Further, to have adequate stellar density for tracing magnetic field at sub-parsec scales means the angular sampling must be greater than $10-20$ reliable sources per square arcmin. From the Two Micron All Sky Survey (2MASS) \citep{skrut06} data, to reach the above angular sampling we need to observe magnitude depths of $13-14\,\mathrm{mag}$ in the galactic plane between latitudes $b\pm1^{\circ}$. Hence, to achieve a polarization uncertainty of around 1\% for stars as faint as $13\,\mathrm{mag}$ (e.g.\ Figure~\ref{fig13}) requires per image exposure times ranging from $30\,\mathrm{s}$ to $40\,\mathrm{s}$ in $H$-band with POLICAN. 

\subsection{Signal-to-noise ratio goals}
The accuracy of polarimetric measurements also depends on the source signal-to-noise ratio (S/N). 2MASS data reached $15\,\mathrm{mag}$ at a S/N of 10 in $H$-band. We aim to obtain a $\mathrm{S/N}>10$ for $15\,\mathrm{mag}$ sources in order to reach the best polarimetric accuracy. This can be achieved by combining more number of images acquired per field for each HWP. Hence, we considered typically a minimum of 15~images are required for each HWP angle, leading to a total of 60~images for a given observing field. This leads to a total integration time of around 30 to $40\,\mathrm{mins}$ for a given field. 

Essentially, these values and estimates form the basis for starlight polarimetry with POLICAN. However, observations of extragalactic and other sources can be customized to have different integration times as desired.

\section{Observation scheme}\label{obscheme}

Ground-based images obtained in the NIR are contaminated by the atmospheric sky emission (OH line and thermal continuum emission), sky transmission noise, and thermal emission from the telescope and optics. Successfully isolating these effects during data processing is essential to consider in an observing scheme. A standard practice for observing in the NIR is to obtain multiple images using telescope dither. This facilitates estimating the aforementioned ``sky'' contributions. Typically, a minimum of five dithered images is sufficient to establish the sky image. However, as noted in the previous section, we want to obtain more images per field to boost the S/N of the combined image. Hence, with POLICAN, we implemented a sequence to acquire 15 dithered images for each HWP angle for a given observing field.

The Stokes parameters are calculated by the difference in flux between two orthogonal polarization measurements (see Section~\ref{polanal}). The flux difference is more accurate when the two orthogonal polarizations are obtained in sequence, as the change in sky transmission is minimum \citep{clemens12a}. Based on this, we designed the observing scheme such that the HWP images corresponding to each Stokes parameter are acquired consecutively for each dithered position (i.e.\ in the order $0^{\circ}$ and $45^{\circ}$ for Stokes $Q$, $22$.$5^{\circ}$ and $67$.$5^{\circ}$ for Stokes $U$). This sequence is followed for all the 15 dither positions leading to a total of 60 images per observed field.

\begin{figure*}[!ht]
\epsscale{1.1}
\plotone{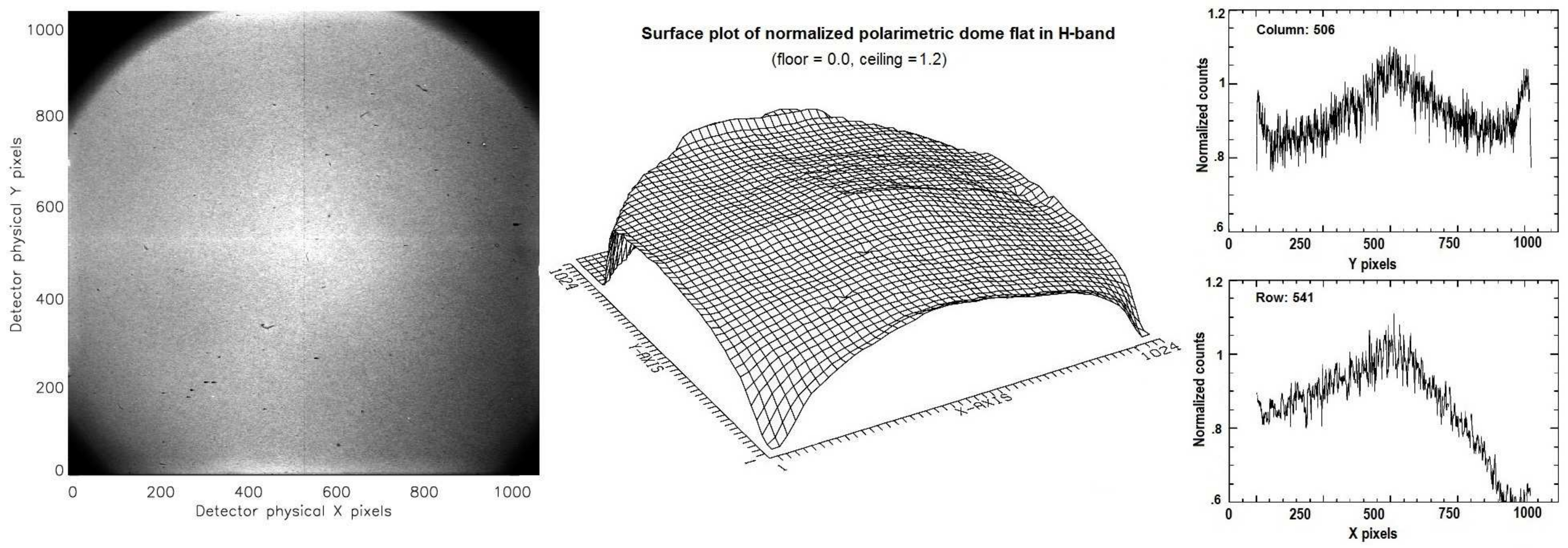} 
\caption{Images displaying various characteristics of a normalized master flat image. The \textit{first panel} shows the flat image marked with detector pixel values in physical coordinates. The dark corners in the image are formed due to vignetting. The \textit{center panel} shows a 3D surface plot of the normalized flat image that reveals the illumination profile and some large-sale aspects of pixel-to-pixel variations. The \textit{last panel} shows row and column cuts through a particular region of pixels across the flat image.}
\label{fig3}
\end{figure*}

The dithering strategy varies depending on the field of interest and size of the source.\\
\indent 1) For studying magnetic field properties through starlight polarimetry, the sequence of observations consists of 15 dithers distributed in a non-repetitive random pattern within a diameter of $30\,\mathrm{arcsec}$ around the targeted center. The dither size of $30\,\mathrm{arcsec}$ makes sure that the extended emission from bright stars do not overlap in each image.\\
\indent 2) For studying scattered polarization from extended sources, the sequence of observation consists of a dithering pattern of 8 source images and 7 off-field images. The source and off-field images are obtained alternatively. The off-field images are taken by dithering completely outside the source field in one or more different cardinal directions (typically along the north-south direction).

To obtain the dark contribution, a set of 10 dark images are acquired at the end of the observing night for all values of exposure time used. The 10 darks for each exposure are then averaged to obtain mean dark images. 

The automatic image acquisition sequence for observations is passed to the telescope control system by a JavaScript from the main computer operated by the observer. The script includes functions for telescope dither, camera exposure time, and HWP rotation. The user can modify each parameter as desired. Once the script is run, user intervention is not needed. A log of the POLICAN observing runs conducted to date is saved both in hard copy and in digital format. On a typical night, POLICAN produces around 600 images, which approximately sum up to $2.5\,\mathrm{GB}$ of memory. 

\subsection{Flat fielding strategy} \label{flat}

There are both temporal and spatial variations in the detector that affect the quality of the images. Temporal variations from noise and sky amplify with spatial variations introduced by non-uniform illumination and pixel-to-pixel variations in the detector. Other effects include dust on the HWP, which rotates during modulation and is not canceled in the Stokes combination scheme. To correct for all of the above effects, a suitable flat-fielding strategy is required. Because polarimetric observations vary with each HWP angle, the flats should be obtained at the same HWP angles.

Based on Mimir calibration \citep{clemens12b} methods, we implemented a similar technique for obtaining these ``polarimetric flats.'' The flats consisted of sequence of images ($\sim30-$empirical estimation) acquired with dome lights ON and OFF for each HWP angle. Exposure times were set so as to fill half of the full-well-depth in the ON images. The lights ON and OFF images are differenced to eliminate the effects of dark counts and thermal backgrounds. The differenced flats are then averaged to obtain a master flat with high S/N. This is repeated for each HWP angle to obtain four master flats. The master flats are normalized using the mean value obtained from a region of 40 $\times$ 40~pixels in the central zone [460:500,~540:580] of the detector. Figure~\ref{fig3} shows different characteristics of a normalized master flat. The center panel shows the 3D surface plot of the normalized master flat where the illumination profile is seen along with large-scale non-uniform pixel-to-pixel variations. The last panel shows row and column cuts obtained for a particular region of pixels. 

\section{Data Processing}\label{dataproc}

The images obtained with POLICAN contain a number of instrumental effects in addition to the NIR atmospheric contamination, leading to a challenge to obtain high-quality, linear polarimetric data. These include 1) the non-linear response of the detector array pixels; 2) the dark current that is pixel, time, and temperature dependent; 3) the bad pixels, about 0.2\% spread across the detector array; 4) the non-uniform illumination profile and the pixel-to-pixel variations; 5) detector crosstalk effects, seen as residual negative charges in different quadrants; 6) the atmospheric sky emission and transmission that varies with time and position; and 7) the thermal emission from telescope and optics that is FOV and time-dependent.

To extract useful polarimetric information from POLICAN observations, all the above problems need to be canceled or mitigated. Hence, a robust data processing method was developed to obtain accurate values of degree of polarization and position angles for each observing field while also managing all necessary corrections. The procedure involves two distinct stages of data processing. The first stage consists of the basic steps of reduction and image correction leading to science-quality images. The second stage utilizes the multiple measurements and analyses from the reduced images to yield degree of polarization, position angles, and their corresponding uncertainties.

\subsection{Basic data processing} \label{BDR}

Stage~1: The first stage of data processing that will lead to science-quality images uses a custom pipeline called \textit{POLREDUCE}, developed in the IRAF\footnote{Image Reduction and Analysis Facility (IRAF) is distributed by the National Optical Astronomy Observatory, which is operated by the Association of Universities for Research in Astronomy (AURA) under a cooperative agreement with the National Science Foundation. \url{http://iraf.noao.edu/}} environment. Image display and interactive functions are carried out with the help of SAO DS9 software. The individual dithered images obtained for a given set of observation are associated based on their HWP angle and filter band. The corresponding mean dark images and polarimetric flats are grouped in the same set. The reduction process is carried out separately for each HWP angle leading to four science-quality images. The steps in reduction are as follows:
 
\indent 1) The first step in reduction involves non-linearity correction. The correction coefficients are estimated by characterizing the illumination response of each pixel using dome flats as described in Paper~I. Using the correction coefficients, the individual images are corrected for each pixel to establish linearity corrected images.\\
\indent 2) Next, the images are subtracted with the mean dark image to remove the dark count contribution. \\
\indent 3) The non-uniform illumination, pixel-to-pixel variations, and any effects introduced by the HWP are corrected by flat fielding using polarimetric flats (see Section~\ref{flat}). \\
\indent 4) The individual dithered images after Step~3 are stacked and median combined using the \textit{imcomb} task to obtain the sky image. The images are then subtracted with the sky image to obtain the reduced ``clean'' images. The reduced images have an offset in their zero level introduced by the median sky. \\
\indent A modal filter is applied to each image with a $3\sigma$ lower and upper threshold. The modal value gives the zero level offset, which is then subtracted from the reduced images, resulting in uniform background nearly to zero. The images are next corrected for detector crosstalk effects as described in Paper~I.\\
\indent Step~4 is slightly modified if the reduction is carried out for extended sources. Because dithering is performed as alternating source and off-field, only the off-field images are median combined to obtain the sky image. The rest of the procedure remains the same.\\
\indent 5) Because the images are shifted because of dithering, they have to be aligned before combining. The aligning procedure is carried out by selecting the centroids ($Xcen$ and $Ycen$) of a common star in all of the images. Using the centroids, the shift in each image, both in $X$~and~$Y$ directions, is computed, keeping the first image as reference. From the computed image shifts, the dithered images are aligned using the \textit{imalign} task. The aligned images are next average combined with a minmax rejection to produce the final reduced image.\\
\indent Image aligning and combination remains the same for extended sources with only the source images used in the process.\\ 
\indent 6) The final corrected image is then transformed to the equatorial system with north-up and east-left direction. Next, the image is cropped to the central 4$\times$4 arcmin$^2$, which removes major optical aberrations (as shown in Paper~I).\\
\indent 7) The last step in the pipeline involves finding all of the point sources using the \textit{daofind} task, with a detection threshold of $5\sigma$. The source positions as centroids $Xcen$ and $Ycen$ are stored in a file. Additionally, all the important parameters involved in the reduction process are saved and can be used in future for rapid re-processing.\\
\indent Steps 1 to 7 are performed for all four sets of images corresponding to each HWP angle, obtaining the final reduced images and source positions. The reference image used in aligning all four sets of images is the same, and hence the final four reduced images are matched in the image coordinate system. The reduced images and results form the basis for polarimetric analysis, which is explained in Section~\ref{polanal}.

\subsubsection{Astrometry corrections}

Observations made with POLICAN have astrometric information updated to the image headers based on telescope and dithering data. These values were found to differ by a few arcsec to a few arcmin when compared with astrometric-based coordinates. The images were also found to have rotation offsets of a few degrees and to possess slight geometric distortions. Hence, scientific analysis of the reduced images required accurate astrometric corrections. Existing astrometry software such as \url{astrometry.net} \citep{lang10} failed to obtain correct solutions, due large offsets in POLICAN image headers, leading to the necessity of implementing a customized program. This is carried out in two steps, both relying on the astrometric information provided by the publicly available 2MASS data.

The first step involves making coarse corrections to center the images such that the astrometric errors are withing a few arcsec. A reference star is chosen in the image, and its corresponding 2MASS coordinates are obtained. The image header is then updated for the central reference value (CRVAL) with the 2MASS coordinates of the reference star. This transforms the image field with astrometric information close to the true coordinates. Users satisfied with this information have the choice to skip the second step and move directly to polarimetric analysis. Most of the time this is the case for observations of extended sources. However, starlight polarimetry requires further corrections. 

The second step involves obtaining solutions to rectify the image rotation and geometric distortions. This is performed with the help of tasks in IRAF \textit{imcoords} package. As the images at this stage have a roughly acceptable astrometry, the 2MASS catalog for all the point sources in the field are obtained within a given search radius. Next, the $Xcen$ and $Ycen$ for a minimum of six point sources in the image are associated with their 2MASS coordinates. These are then used to obtain the plate solutions in the equatorial system. The astrometry correction for one HWP image is sufficient to correct the other 3 HWP images, as they are aligned with respect to each other. The astrometry information of the first image is directly copied to the other three, making them equal both in image and celestial coordinates. 

\subsection{Polarimetric Analysis} \label{polanal}

Stage~2: The second stage of polarimetric data processing is carried out using a pipeline developed in Interactive Data Language (IDL). The pipeline is called \textit{FLX2POL}, and it combines functions and procedures that are found in the IDL Astronomy Library \citep{land93} and in the Coyote Graphics Library\footnote{\url{http://www.idlcoyote.com/}}. The main steps in polarimetric analysis are to accurately measure the flux of sources; establish the Stokes parameters; and compute polarizations, position angles, and their corresponding uncertainties. 

The output file from Stage~1 consists of source positions for all point sources. The first step is to match source positions among the four HWP images and obtain a common identification and location list. The source-matching algorithm selects the centroids $Xcen$ and $Ycen$ for all sources in the four HWP images and runs a cross-correlation search within a given radius, typically 2~pixels ($\sim0.{\arcsec}6$). Once the sources are matched, they are sorted in sequence and given a common identification number.

Polarimetric analysis of point source involves measuring the total flux in each of the four HWP images, which is mainly achieved by synthetic aperture photometry. In POLICAN pipeline, synthetic aperture photometry is performed using the \textit{aper} routine adapted from the DAOPHOT \citep{stet87} package. The pipeline includes two runs of photometry that allow aperture selection for obtaining accurate flux values. The first run of photometry involves calculating the S/N of the source with an optimum aperture radius. In the second run, the final flux of the source is measured with an aperture whose value is chosen depending on the S/N of the source. The necessity of aperture selection based on S/N is required, as the PSF of the sources vary both in time and across the FOV as shown in Paper~I. 

\begin{deluxetable}{cc}
\tablecolumns{2}
\tablewidth{0pc}
\tablecaption{Aperture ranges used for Photometry.\label{tbl-1}}
\tablehead{\colhead{Source S/N} & \colhead{Photometric Aperture} \\
\colhead{at 10 pixels} & \colhead{Radius (pixels)}}
\startdata
$<$ 10 			& 7 	\\		
10 - 50 			& 8 	\\
50 - 100 		& 9 	\\
100 - 500 		& 10 	\\
500 - 1000 		& 11 	\\
1000 - 5000 		& 12 	\\
5000 - 10000 	& 13 	\\
$>$ 10000 		& 14 	\\						
\enddata
\end{deluxetable}

Based on magnitude growth analysis \citep{howell89,stet90}, we find that an aperture radius of 10~pixels is optimum for photometry. Using this radius, we perform first run of photometry and obtain the S/N of all the sources. Next, a new aperture is chosen between a range of 7 to 14~pixels (empirical estimation) for each source depending on its S/N (see Table~\ref{tbl-1}). Using the new aperture, a second run of photometry is performed on each source, obtaining the final flux and flux error values which are used to establish the Stokes parameters and their uncertainties. The final flux values in each of the HWP image can be denoted as $I_{0}, I_{22.5}, I_{45}, I_{67.5}$.

The Stokes $I$, $Q$, and $U$ in the instrumental reference system are now computed as follows:
\begin{equation} \label{eqn1}
I = (I_{0}+I_{45}+I_{22.5}+I_{67.5})/2
\end{equation}
\begin{equation} 
Q = (I_{0}-I_{45})/I
\end{equation}
\begin{equation} 
U = (I_{22.5}-I_{67.5})/I
\end{equation}

These are then scaled by polarization efficiency $\eta$ and rotated by the HWP zero-phase offset angle, $\theta$ (see Section~\ref{pcalstnd}) to obtain the equatorial Stokes values. The polarization efficiency $\eta$ was obtained from SIRPOL measurements \citep{kandori06}, as both instruments use the same polarizing elements from the same manufacturers. The $\eta$ values are $J=0.955$, $H=0.963$, and $K=0.985$. The calculation is as follows:
\begin{equation}
Q_{eq} = (Q cos(2\theta) - U sin(2\theta))/\eta \label{eqn4}
\end{equation}
\begin{equation}
U_{eq} = (U cos(2\theta) + Q sin(2\theta))/\eta \label{eqn5}
\end{equation}

Next, the Stokes values are corrected for instrumental polarization, derived from globular cluster observations (see Section~\ref{pcalgc}):
\begin{equation}
Q_c = Q_{eq} - Q_{inst} \label{eqn6}
\end{equation}
\begin{equation}
U_c = U_{eq} - U_{inst} \label{eqn7}
\end{equation}

Finally, the corrected Stokes values are combined to form the equatorial degree of polarization, $P_{eq}$, and the position angles, $P.A$, measured from the north-up to east-left direction:

\begin{equation}
P_{eq} = 100\sqrt{Q_{c}^{2}+U_{c}^{2}} \label{eqn8}
\end{equation}
\begin{equation}
P.A = \frac{1}{2}\tan^{-1}\left(\frac{U_{c}}{Q_{c}}\right). \label{eqn9}
\end{equation}

The $P.A$ values have an ambiguity in the calculation because the arctangent function produces values between $-\pi/2$ to $\pi/2$ radians. They are corrected to represent them within the range of 0 to 180$^{\circ}$. The polarization uncertainty $\sigma_{P}$ is computed from the corresponding Stokes errors which is described in Appendix~\ref{polerr}. The calculated polarization values have a positive bias because of the quadrature combination of the Stokes parameters. The Ricean correction prescription of \citet{wardle74} which works well for polarization S/N values greater than 2 (i.e.\ $P/\sigma_{P} \ge 2)$ is used for de-biasing the polarization values as follows:

\begin{equation}
P = \sqrt{P_{eq}^{2}- \sigma_{P}^{2}} \label{eqn10}
\end{equation}

Photometric measurements for each source is next obtained from the deep co-added intensity image calculated from the four HWP images as in equation~\ref{eqn1}. Based on the previously obtained image source list, aperture photometry is performed on all the sources to obtain their magnitude values. A broad zeropoint correction on instrumental magnitudes is applied by using the average zeropoint value. The source coordinates are converted from pixel values to celestial coordinates using the \textit{xyad} routine, which uses the astrometric information available in the image header. 

The values of polarimetric analysis together with astrometric and photometric measurements are combined to form a catalog of results. The catalog contains the source information as follows: \\
$ID$, $RA$, $DEC$, $Xcen$, $Ycen$, $P\%$, $Perr$, $P.A$, $P.Aerr$, $Mag$, $MagErr$, $Q\%$, $U\%$, $Qerr$, $Uerr$

Polarization visualization are carried out by producing map of vectors representing $P$ and $P.A$ for each source using a separate customized program (e.g~Figure~\ref{fig14}).

Polarimetric analysis for extended sources remain similar to the above description. Because the four HWP images are aligned with each other, the surface brightness in each pixel is used for analysis instead of flux of a point source. The Stokes parameters and polarization values are obtained in order up to equation~\ref{eqn10} for each pixel. The results are then stored into an image array. The visualization and map making procedure is carried out by binning the polarization values in each pixel, with a certain threshold as required by user. For each bin, a vector is plotted to represent the $P$ and $P.A$ value (e.g~Figure~\ref{fig8}).
 
\begin{figure}[!ht]
\epsscale{1.2}
\plotone{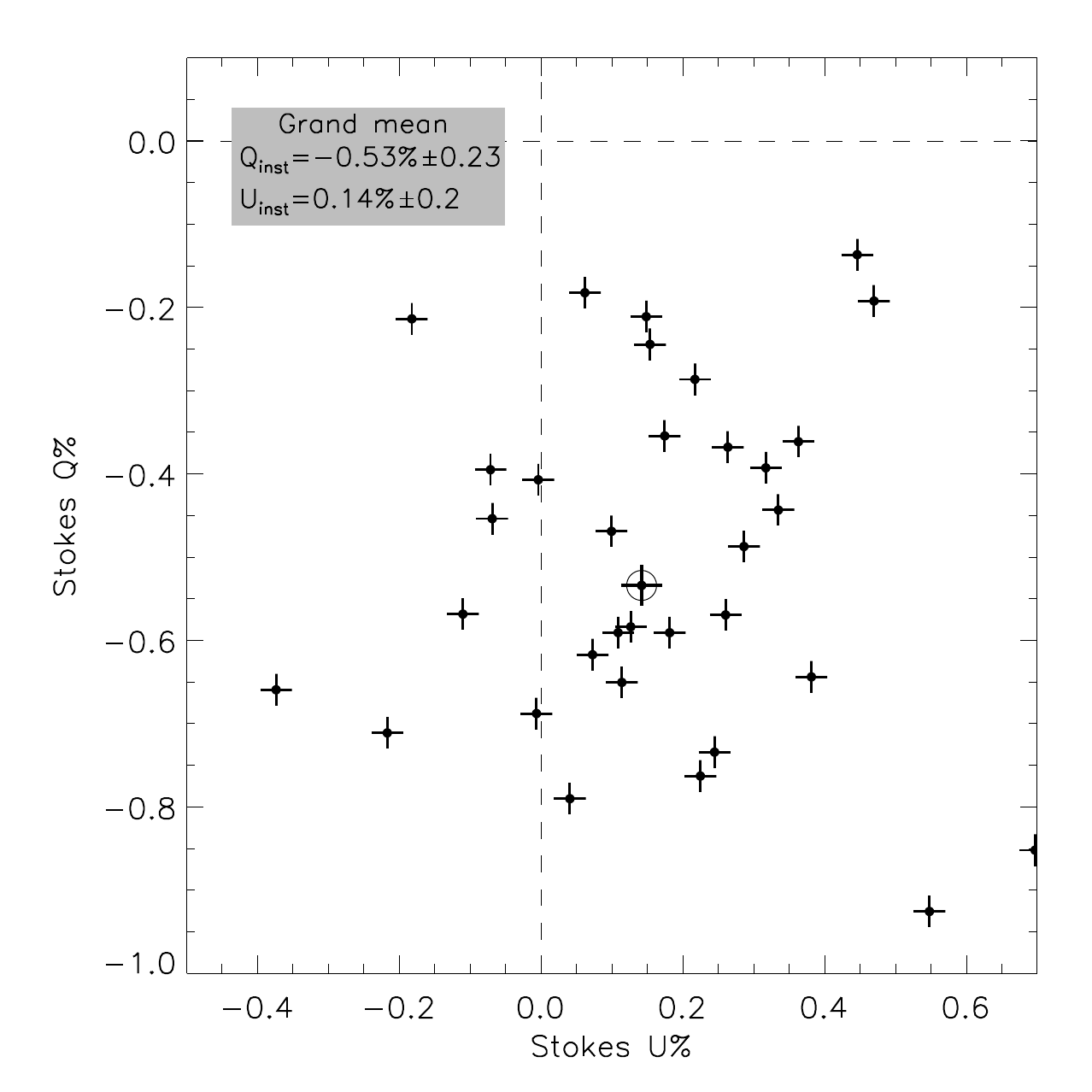} 
 \caption{ Distribution of mean instrumental Stokes $\langle Q\rangle$ and $\langle U\rangle$ values from 37 observations of globular cluster M5. The mean Stokes values for each observation are represented by plus symbol. The final grand mean of all the 37 mean Stokes values is represented by a circle with plus symbol. }
\label{fig4}
\end{figure}

\section{Polarimetric Calibration}

Design factors and optical setup introduce instrumental polarization that need to be carefully removed to faithfully recover the true polarization. As the polarizing elements in POLICAN are located ahead of CANICA, the only contribution for instrumental polarization should be the telescope mirrors. POLICAN is designed for single-beam linear polarimetric observations with just two polarizing components: the HWP and polarizer. There exists no room for other optical elements such as prisms in the setup for producing artificial polarization for calibrations. Hence, all the calibration needs to be done through observations of astronomical objects. For POLICAN, we used two standard steps of polarimetric calibration as described for the Mimir instrument \citep{clemens12b}: 1) removal of instrumental polarization across the FOV from observations of globular clusters and 2) converting instrumental polarization position angle to equatorial position angle from observations of polarimetric standards. In the following sections, we describe the methods and results obtained for polarimetric calibration. 

\begin{figure*}[!ht]
\epsscale{1.0}
\plotone{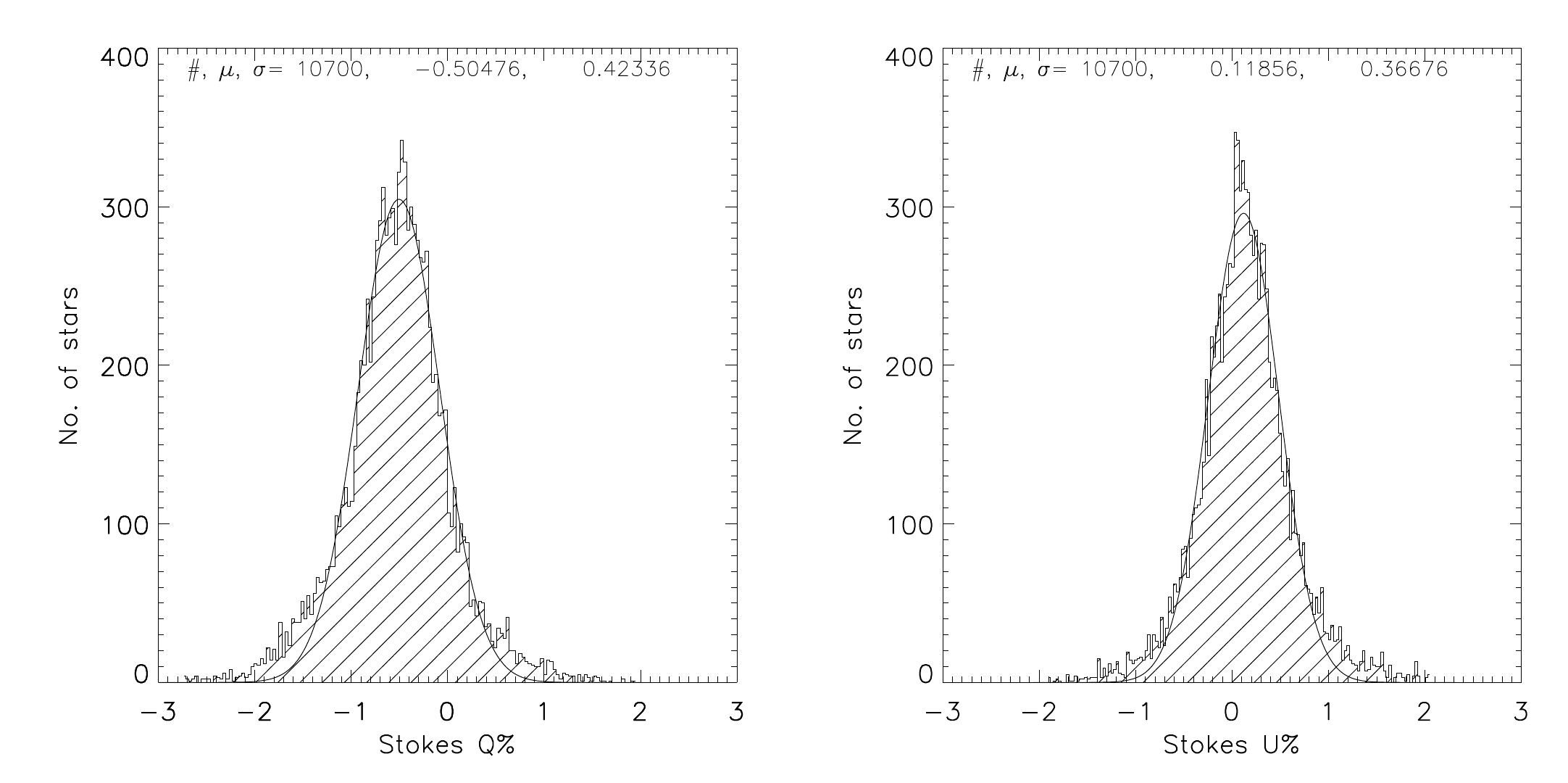} 
 \caption{Histogram of instrumental Stokes $Q$\% and $U$\% for combined values of 10,700 stars in the 37 observations of globular cluster M5. The histograms are fitted with a Gaussian to calculate the mean and standard deviation of the distribution. The combined data of all the Stokes values produced the final instrumental polarization for POLICAN as $Q_{inst}=-0.50\%$, $U_{inst}=0.11\%$ with $P_{inst}=0.51\%$.}
\label{fig5}
\end{figure*}

\subsection{Globular Clusters} \label{pcalgc}

Globular clusters are known to have stars with low polarization levels. Clusters with large angular extent are useful to calculate the instrumental polarization across the entire FOV. Mimir calibration observations of multiple globular clusters (M2, M3, M5, M12, M13 $\mathrm{etc.}$) showed that \objectname{M5} has polarization values below $0.1\%$ with low color excess $E(B-V)$, representing the best cluster with distributed stars. Hence, we concentrated on observing only \objectname{M5} to determine the instrumental polarization for POLICAN. M5 observations were conducted in various runs over a period of three years from 2013 to 2016. In total 37 sets of clear sky observations were obtained, which was used for analysis. The images were acquired using dithering methodology for 15 positions as described for extended sources in Section~\ref{obscheme}. In all observing runs, the image acquisition order remained same, with exposure time set to $20\,\mathrm{s}$. The M5 fields were distributed to fall on the detector center as well as at different quadrants of the detector. This allowed to have distributed polarization values in all the pixels, making it feasible to map the instrumental polarization across the entire FOV.  

The images from each observation were reduced as described in Section~\ref{BDR}. All the stars in each field were selected and analyzed up to calculations of Stokes values in equatorial system (i.e.\ up to equation~\ref{eqn5}), as described in Section~\ref{polanal}. By the time of analysis, the HWP zero-phase offset angle $\theta$ was determined and hence we converted the Stokes values into equatorial system. Next, the polarimetric and photometric values of the stars were obtained up to equation~\ref{eqn10}, omitting equation~\ref{eqn6}~and~\ref{eqn7}. This produced a catalog of results for thousands of stars in each observing field. The results were filtered to obtain Stokes values only for bright stars with low polarimetric uncertainties. (i.e.\ for stars with $\mathrm{mag}<13$ and $\sigma_{P}<1.0$). This avoided the faint stars with larger errors and forced the measurements to true instrumental polarization. Further, the central crowded field of stars were excluded, because aperture photometry fails to obtain accurate measurements in crowded regions. An alternative to aperture photometry is PSF photometry, but this could not be applied, as CANICA images have varying PSF across the FOV as described in Paper~I.

Next, for each observation, the mean Stokes $\langle Q\rangle$ and $\langle U\rangle$ were calculated from the individual Stokes values of the selected sample of stars. These were then combined to form the grand mean Stokes value $\langle Q_{inst}\rangle$ and $\langle U_{inst}\rangle$. Figure~\ref{fig4} shows the distribution of mean Stokes $\langle Q\rangle$ and $\langle U\rangle$ in plus symbol for the 37 observations of M5. The grand mean Stokes value is shown at the center with a circle and plus symbol. The estimated values of grand mean instrumental Stokes values are $\langle Q_{inst}\rangle=-0.53\%\pm0.23$, $\langle U_{inst}\rangle=0.14\%\pm0.2$. 

To obtain the final value of instrumental polarization across the FOV, we combined results of all the selected sample of stars in the 37 observations. A total of 10,700 stars were obtained in the combined result. The combined Stokes $Q$ and $U$ values for all the stars were examined by a histogram and a Gaussian\footnote{In each plot of Gaussian fit, \# indicates the total number of measurements in the distribution, $\mu$ indicates the peak value of the Gaussian fit, and $\sigma$ indicates the standard deviation of the Gaussian fit.} was fitted to the distribution as shown in Figure~\ref{fig5}. The peak of the histogram matched with the peak of the Gaussian fit, which represented the final instrumental Stokes value of POLICAN. The values were estimated to be $Q_{inst}=-0.50\%$ and $U_{inst}=0.12\%$. These values remained consistent within $0.02\%$ levels when compared with the grand mean Stokes value. The uncertainties in the instrumental Stokes $\sigma_{Qinst}$ and $\sigma_{Uinst}$ were taken as the standard deviation of the fit, which gave $\sigma_{Qinst}=0.42\%$ and $\sigma_{Uinst}=0.36\%$.

\begin{figure}[!ht]
\epsscale{0.92}
\plotone{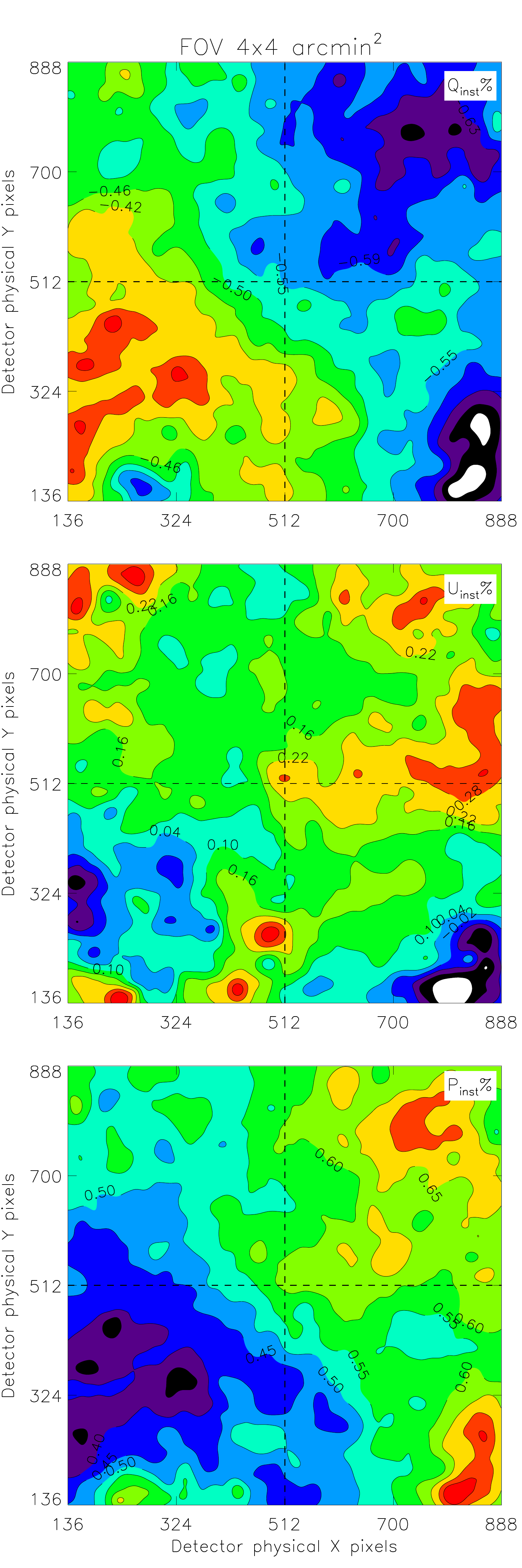} 
\caption{POLICAN instrumental polarization map of Stokes $Q_{inst}$, $U_{inst}$ and $P_{inst}$ across the central 4$\times$4 arcmin$^2$ FOV, obtained from observations of globular cluster M5. The contours for $Q_{inst}$, $U_{inst}$ and $P_{inst}$ start at -0.7\%, -0.2\%, 0.3\% and are stepped in increasing order of 0.04\%, 0.06\% and 0.05\% values up to 10 contour levels.}
\label{fig6}
\end{figure}

The full-field instrumental Stokes values were next calculated from the same sample of 10,700 stars. The Stokes values for all the stars were sorted by their positions and gridded into an image array with resolution of 1~pixel. The image arrays were next smoothed to 10~pixels and converted into a contour map to represent the Stokes value for a FOV of 4$\times$4 arcmin$^2$. Figure~\ref{fig6} shows the map of instrumental Stokes $Q_{inst}$ and $U_{inst}$, along with the $P_{inst}$ value computed from the former Stokes values. The variations in instrumental Stokes, both $Q$ and $U$, across the FOV are around $\pm0.2\%$. This variation is below the Stokes uncertainties, indicating FOV dependence of instrumental polarization is minimum. In summary, the final instrumental Stokes $Q_{inst}$ and $U_{inst}$ are calculated to be $-0.50\%$ and $0.12\%$, and the mean instrumental polarization is $0.51\%$.

\subsection{Polarimetric Standards} \label{pcalstnd}

POLICAN's mechanical assembly and the polarizing elements are fixed along their axis and mounted stationarily to the telescope. It is not definitive that the HWP zero-phase angle is aligned to the equatorial north. This results in calculations of polarization position angle to be based in the instrumental coordinates. Hence, it is important to determine the HWP zero-phase offset, to correct the position angle to standard equatorial system. Observation of polarimetric standards are the best way to determine the offset angle. Mimir's calibration observations of polarimetric standards provided a large sample of stars for the study. They were mainly derived from \citet{whittet92}, who studied wavelength dependence of polarization. The $H$-band filter central wavelength and bandwidths remained the same for POLICAN and Mimir, which avoided any corrections for wavelength dependence.  

A large number of calibration observations during each run were directed towards two bright standard stars that are available during most times of the year. These were \objectname{HD38563C} and \objectname{CygnusOB221} in the fields of Orion and Cygnus. Here, we present results obtained for HD38563C which was observed for a total of 19~nights during a period of 6~months. 

\begin{figure}[!ht]
\epsscale{1.25}
\plotone{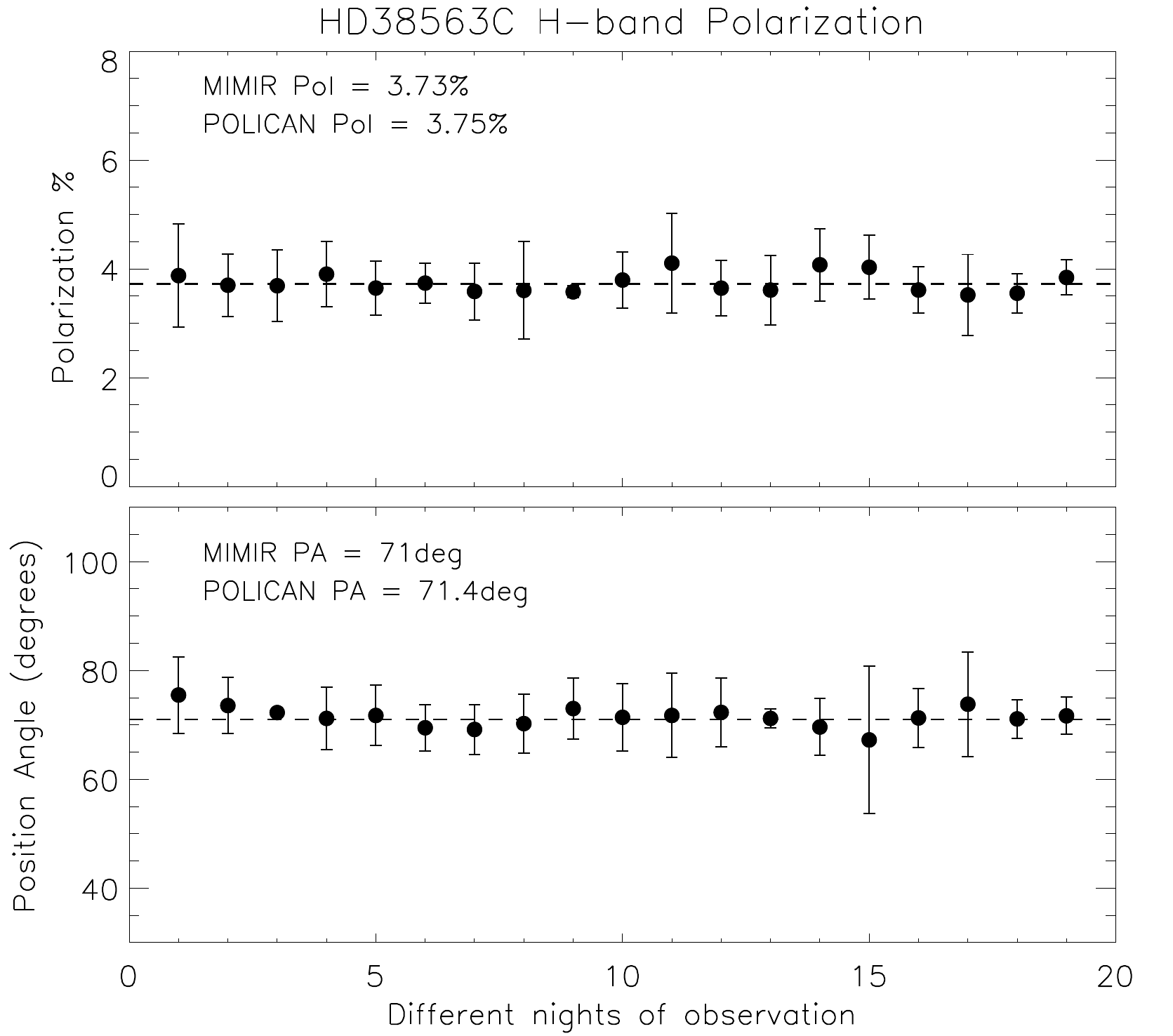} 
\caption{Polarimetric calibration results of standard star HD38653C for 19 different nights of observation. The \textit{top panel} shows distribution of polarization values with their error bars against different nights. The dashed line represents the published value of 3.73\%. The \textit{bottom panel} shows distribution of calculated position angles with their error bars. The position angle values are corrected for HWP zero-phase offset. The dashed line represents the published value of 71$^{\circ}$.}
\label{fig7}
\end{figure}

The observation scheme for HD38563C was similar to point sources as described in Section~\ref{obscheme} with exposure time set to $5\,\mathrm{s}$. The HD38563C field was targeted to fall on the detector center to avoid any effects introduced by optical aberrations. The images were reduced and analyzed up to equation~\ref{eqn6}~and~\ref{eqn7} as described in Section~\ref{polanal}, to obtain the corrected Stokes values $Q_c$ and $U_c$. The only change in the analysis was that the HWP offset angle was set to zero. The instrumental polarization calculated from globular clusters remained in the instrumental system for this analysis.

The individual Stokes values, $Q_c$ and $U_c$ from each of the 19 observations were averaged to obtain the mean Stokes value. Next, the mean Stokes value was used to compute the Ricean corrected polarization and position angle as described in equation~\ref{eqn8}~to~\ref{eqn10}. The computed position angle was then compared with the Mimir's published value to find the difference. The difference gave the HWP zero-phase offset angle for POLICAN, which was determined to be $139^{\circ}$.

Figure~\ref{fig7} shows results of HD38563C corrected for position angle for all the 19 observations. Results from each observing night are represented by a black-filled circle with their corresponding error bars. The final computed mean polarization and position angle are expressed on the top left corner of each panel. Overall observations showed good agreement with published values. Over the course of the last few years, there were times when the motor sensor failed to locate the HWP home position. This was corrected immediately and the sensor was brought back to the original setting to keep the home position constant, in turn the offset angle for POLICAN remained consistent.

\begin{figure*}[!ht]
\epsscale{1.07}
\plotone{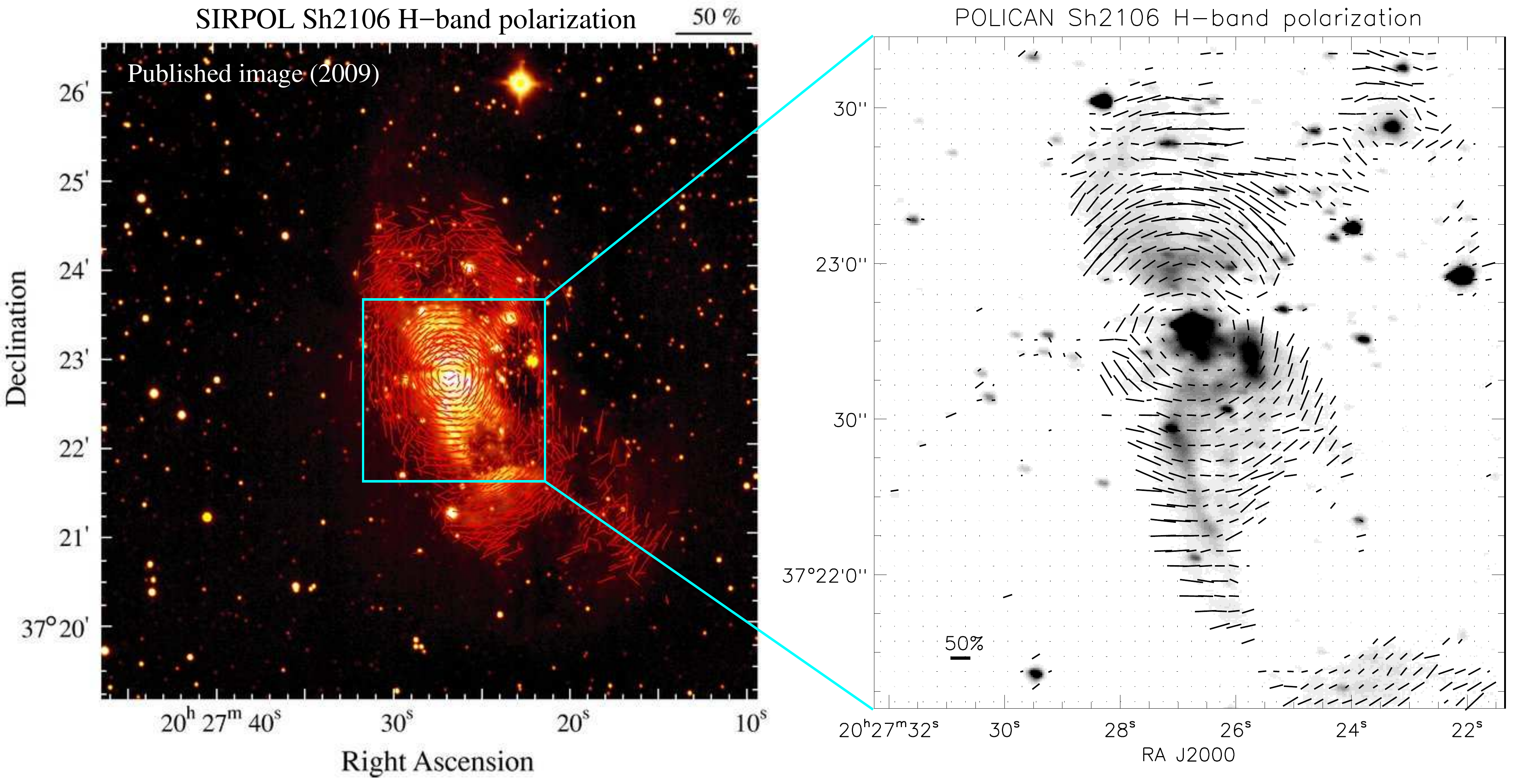} 
\caption{SIRPOL and POLICAN $H$-band polarimetric observations towards Sh 2-106. The \textit{left panel} is a
published image of 7.7$\times$7.7 arcmin$^2$ by \citet{saito09}. The \textit{right panel} shows POLICAN results for FOV of 4$\times$4 arcmin$^2$. The length of the vectors represent the degree of polarization and their orientation represents the position angle. The vectors are plotted only for values above $3\sigma$ for every bin of $3\times3$ pixels. A reference vector at bottom left indicates a value of 50\%.}
\label{fig8}
\end{figure*}

\section{Observations}

POLICAN saw its first light at the end of 2012. Over the past few years, the majority of the observations were concentrated towards polarimetric calibration and pilot studies. These observations were spread evenly in each semester for an average telescope time of around 7 to 8 weeks in a year. Early science proposals were targeted for well known sources to provide comparison of POLICAN results with the literature data. In parallel, a few new regions of interest were also observed to compliment the current studies with priority. These included galactic molecular clouds, H\,{\sc ii} regions, planetary nebulas, and post-AGB stars. 

The starlight polarimetry towards the center of the galactic plane is usually around 3\% polarization due to the presence of large columns of dust. These regions provide a good starting point for pilot studies. The GPIPS survey \citep{clemens12a} spanning from $18^\circ\le l\le 56^\circ$ and $-1^\circ\le b\le 1^\circ$, contains an excellent catalog of polarimetric data for comparing the POLICAN observations. 

To evaluate the performance of POLICAN, we observed two regions that represented good examples of scattered polarization and starlight polarimetry. They were Sharpless H\,{\sc ii} region \object{Sh 2-106} and the GPIPS field number 182. In the following sections, we present the observation details for each of them.

\subsection{Scattered polarization}

\object{Sh 2-106} is an emission nebula estimated to be at a distance of $600\,\mathrm{pc}$ in the Cygnus constellation. At the center of the nebula is a young massive star (type O8) of approximately 15~solar masses, which emits jets of gas forming a bipolar structure. We carried out $H$-band polarimetric observation towards the central 4$\times$4 arcmin$^2$ region surrounding the massive star. The observation scheme was similar to extended sources as described in Section~\ref{obscheme} with exposure time set to $20\,\mathrm{s}$. Image reduction and analysis followed the steps described in Section~\ref{BDR}~and~\ref{polanal}. The final polarimetric results were saved into an image array.

\begin{figure*}[!ht]
\epsscale{1.1}
\plotone{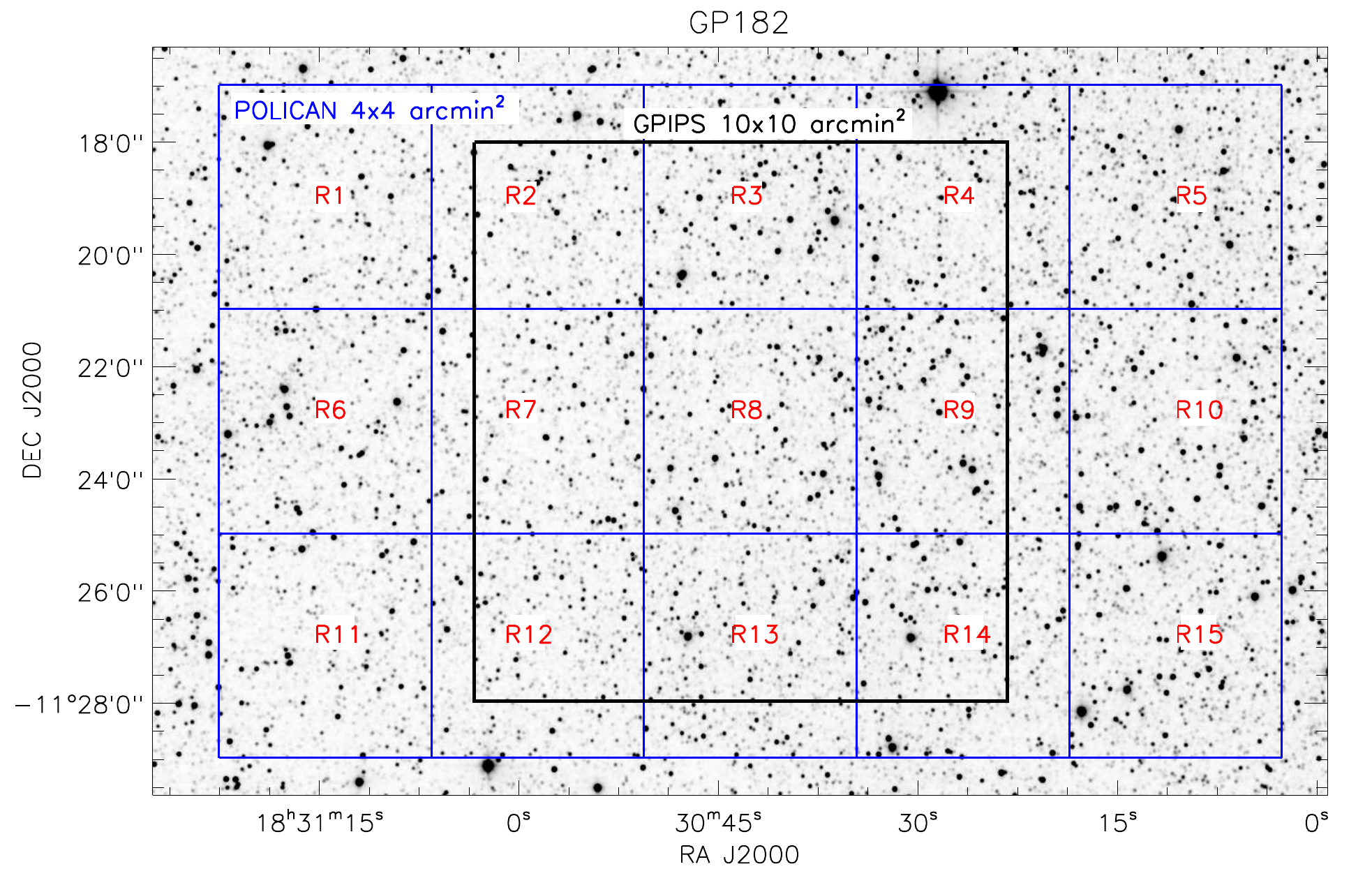} 
\caption{The GPIPS field 182 and its surrounding regions are shown in the background image from 2MASS $H$-band survey. The  $10 \times 10\,\mathrm{arcmin^2}$ FOV of Mimir instrument centering GP182 is shown in black box. The POLICAN pointings for mapping this entire region are shown by blue box with each field marked in its observing order from R1 to R15. The FOV for each POLICAN pointing is $4\times4\,\mathrm{arcmin^2}$.}
\label{fig9}
\end{figure*}

Analysis of the results showed that the region surrounding the central star had high polarization levels to $\sim50\%$. The distribution of position angles appear to be in a centro-symmetric pattern, as had been previously shown by \citet{saito09}. This indicates that the nebulosity is illuminated by the central star which causes the strongly polarized light via dust scattering. Further, the distribution of bipolar structure shows the circumstellar matter associated to the massive star, similar to a disk/envelope system. In Figure~\ref{fig8} we show the results of SIRPOL \citep{saito09} and POLICAN observations of Sh 2-106. The right panel displays POLICAN polarization vectors plotted for surface brightness above 3$\sigma$, binned for every $3\times3$~pixels. Comparing the results in Figure~\ref{fig8}, we see that the angular and spatial resolution achieved with POLICAN is better than SIRPOL. This allows us to resolve dense regions around massive stars to have a clear distinction of nebulosities and circumstellar matter. Such observations with POLICAN will help to obtain scattered polarization at smaller scales making it vital for star-forming studies.

\subsection{Starlight polarization}

The GPIPS field 182 (hereafter GP182) is centered around the galactic coordinates of $l=20.456$ and $b=-0.645$ with a size of $10 \times 10\,\mathrm{arcmin^2}$. \citet{clemens12c} showed that GP182 field contains high stellar density with significant polarization detections. Further, the stars in the field have high polarization S/N ($P_{S/N}=P/\sigma_{P}$) and their galactic position angles are oriented along the galactic plane. Hence, GP182 field is an ideal region for evaluating POLICAN's performance. We chose a total region covering $20 \times 12\,\mathrm{arcmin^2}$ for mapping GP182 and its surrounding areas. The useful FOV with POLICAN is $4 \times 4\,\mathrm{arcmin^2}$, therefore the observations need to span multiple pointings to cover the entire region. By equally placing $4 \times 4\,\mathrm{arcmin^2}$ fields distributed over the entire region, we obtained a total of 15 pointings for POLICAN. Figure~\ref{fig9} shows the background 2MASS image overlaid with 10~arcmin field of GP182 in black color. The 15 POLICAN $4\,\mathrm{arcmin}$ fields are shown in blue color and are marked from R1 to R15, based on their observing orders. The fields are not overlapped with each other, as the full FOV of POLICAN is $5.5 \times 5.5\,\mathrm{arcmin^2}$ and during image reduction they are cropped to $4\,\mathrm{arcmin}$ fields.

GP182 observations with POLICAN were conducted for four nights during 2017 April. Each field was observed with 15 dither positions totaling 60 images for all the HWP angles. The exposure time was fixed to $20\,\mathrm{s}$ with a dither diameter of $30\,\mathrm{arcsec}$. During each night three to four fields were observed and started at the same universal time to keep the airmass and time-dependent variations minimum. The total clock time taken to complete the 15 fields was $7.5\,\mathrm{hours}$. The basic reduction included linearity correction, dark subtraction, polarimetric flat-fielding, sky subtraction, and image combination. Thereduced images were astrometry corrected and processed for polarimetric analysis through aperture photometry. The detailed steps in reduction and analysis followed the description in Section~\ref{BDR} and \ref{polanal}. Once the astrometric, photometric, and polarimetric results were obtained for all the 15 fields, they were combined into a catalog representing POLICAN data for the entire region. The results and comparison to GPIPS data are discussed in the Section~\ref{res}~and~\ref{per}. The mapping strategy by having multiple pointings for large regions forms the basis for polarimetric observations of molecular clouds, filaments and H\,{\sc ii} in the ISM. Given the large clock time for such observations, the regions are limited to sizes within $20 \times 20\,\mathrm{arcmin^2}$.

\begin{figure}[!ht]
\epsscale{1.2}
\plotone{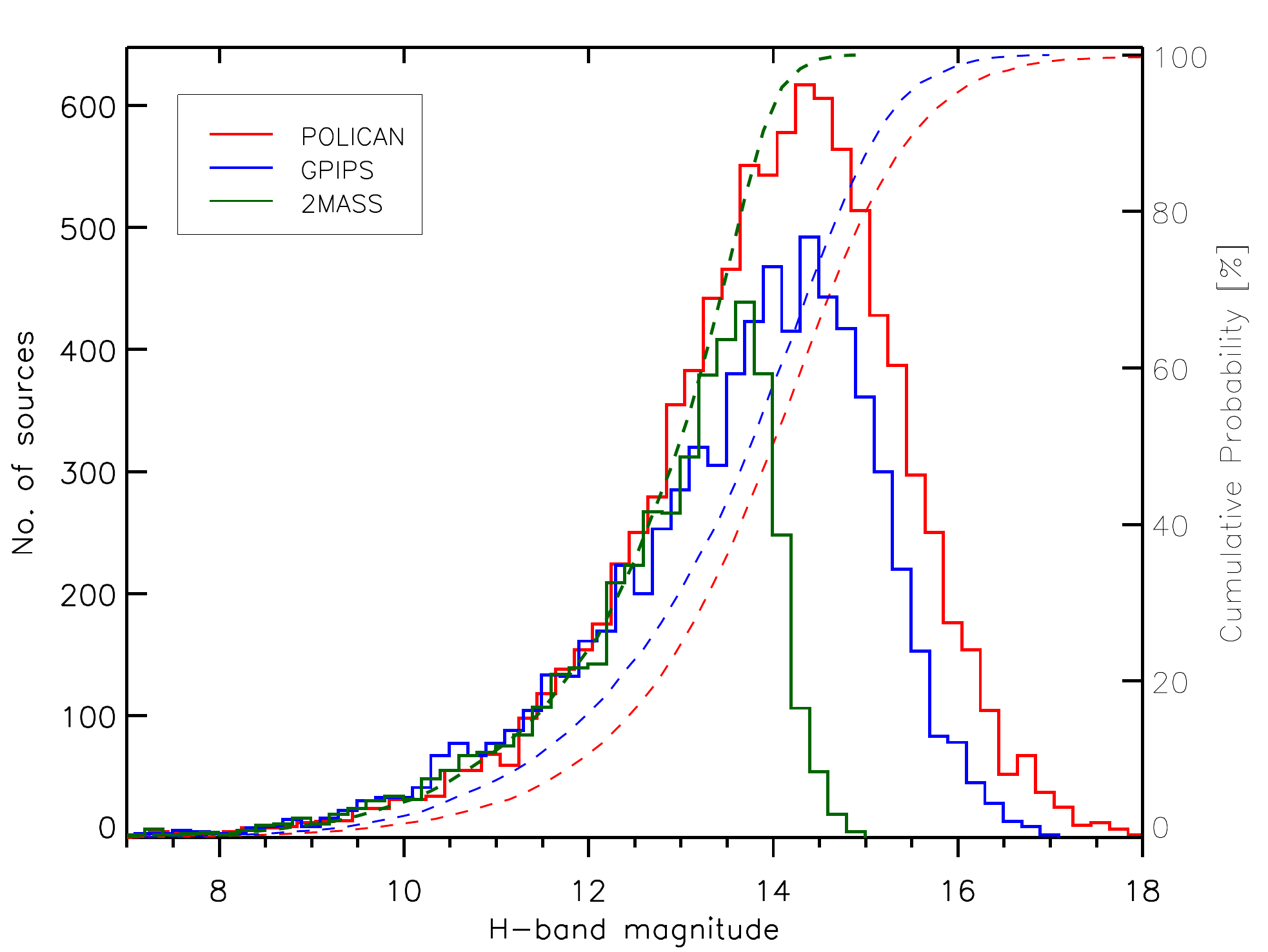} 
\caption{Histogram distribution and cumulative values for POLICAN, GPIPS, and 2MASS stellar detections toward the GP182 region. The horizontal axis gives the $H$-band stellar magnitude. About 50\% of POLICAN detections have stars brighter than $14\,\mathrm{mag}$. The other half detections reach magnitudes as faint as $18\,\mathrm{mag}$.}
\label{fig10}
\end{figure}

\section{Results} \label{res}
\subsection{Stellar properties}

The combined starlight polarimetric catalog towards GP182 region for all the 15 POLICAN fields resulted in a total stellar count of 13,635 stars. These were obtained by selecting sources above $5\sigma$ in the deep co-added intensity image. Out of this entire stellar population, 9556 stars had definite polarization detections. They formed the essential sample of stars for all future analysis. Analyzing 2MASS survey data for the same region, we find the number of stellar count obtained is 4453 stars. This showed that POLICAN observations had twice the number of detections to 2MASS. Similarly, analyzing GPIPS data for the same region, we obtain a total of 7230 stars with definite polarization detections. This indicated the number of polarization detections with POLICAN is much higher for the chosen integration time. Figure~\ref{fig10} shows the stellar count histogram against magnitude for POLICAN, GPIPS and 2MASS data. Also plotted is the cumulative distribution function for each data. It is seen that POLICAN observations reached depth of many orders better than 2MASS. The majority of stars were in the magnitude range from $13\,\mathrm{mag}$ to $16\,\mathrm{mag}$, with 50\% probability of detection for $14\,\mathrm{mag}$ stars. The stellar density achieved with POLICAN in this region is about $30-40$ stars per square arcmin, meeting the sampling goals as described in Section~\ref{obgoals}.

\begin{figure}[!ht]
\epsscale{1.1}
\plotone{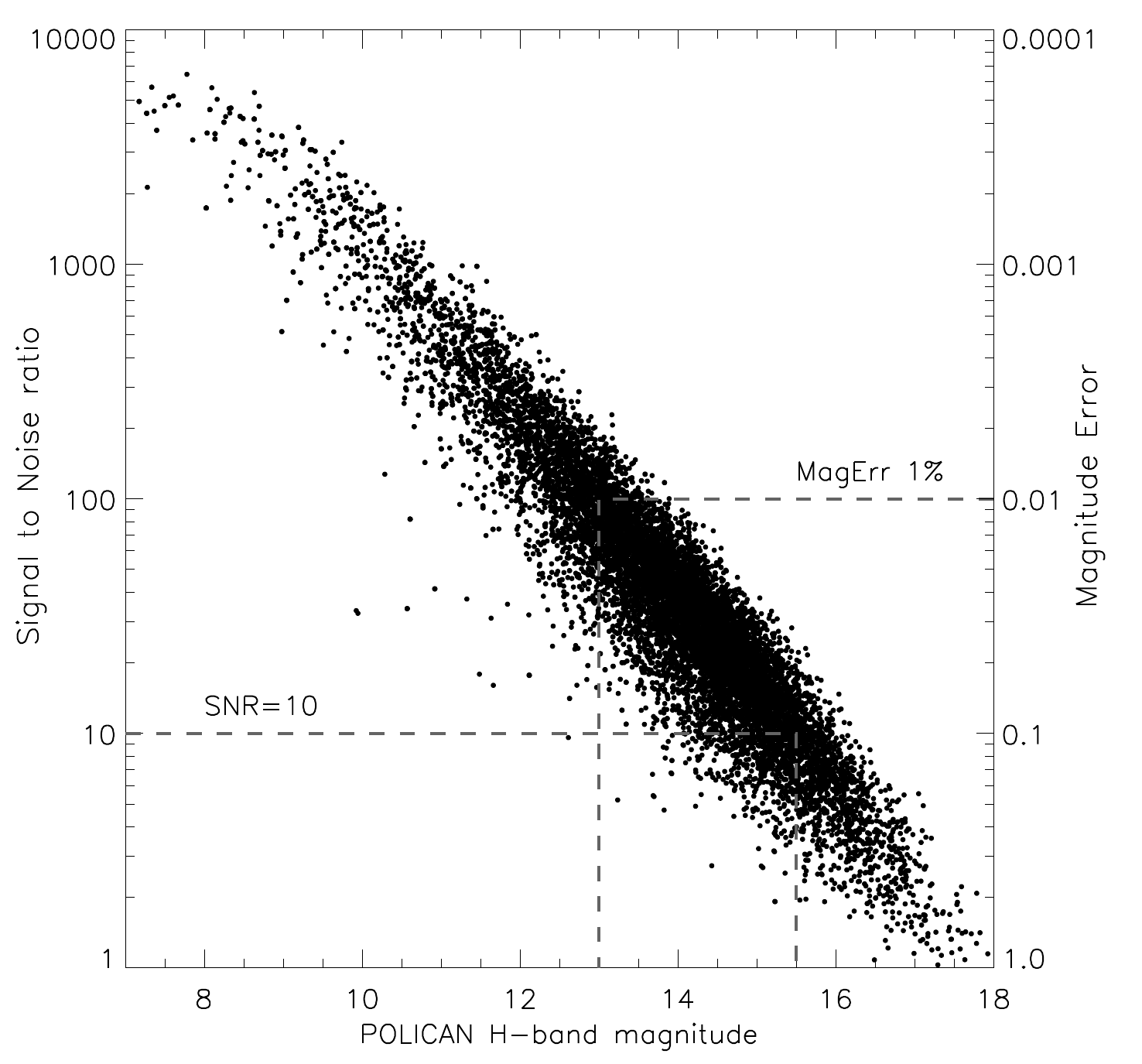} 
\caption{Plot of S/N and magnitude error for all the POLICAN stellar detections towards GP182 region. The horizontal axis gives the POLICAN $H$-band stellar magnitude. The S/N achieved is better than 10 for stars up to $15.5\,\mathrm{mag}$. For stars brighter than  $13\,\mathrm{mag}$, the magnitude errors are better than 1\%.}
\label{fig11} 
\end{figure}

\begin{figure*}[!ht]
\epsscale{1.0}
\plotone{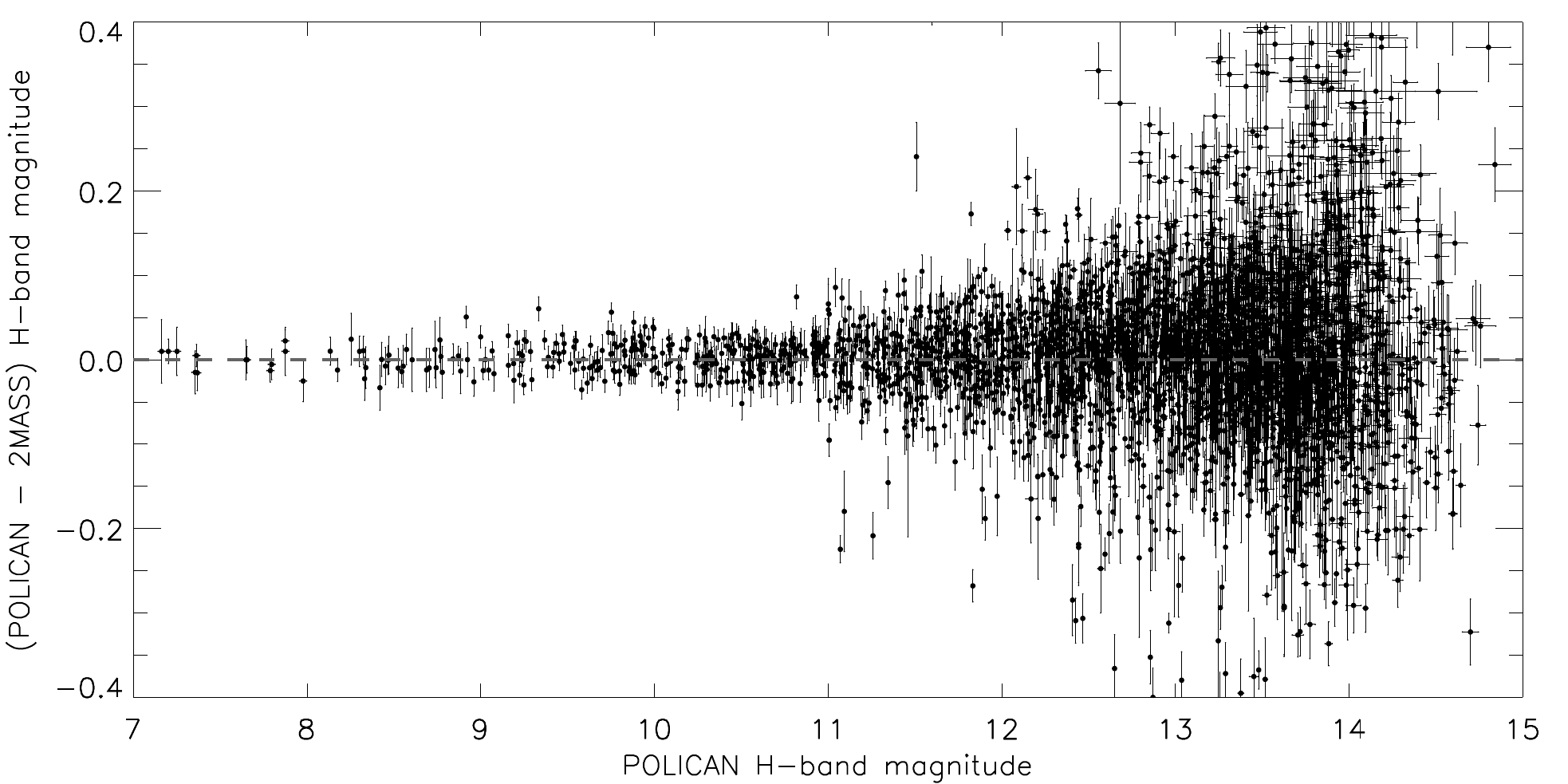} 
\caption{Photometric comparison of POLICAN and 2MASS $H$-band magnitudes for 4453 matched stars towards the GP182 region. The error bars in horizontal and vertical axis represents POLICAN's and 2MASS magnitude errors. The dispersion in magnitude difference is below $0.05\,\mathrm{mag}$ for stars up to $11\,\mathrm{mag}$.}
\label{fig12}
\end{figure*}

\subsection{Photometric properties}

The photometric results obtained from the deep co-added intensity image for the 9556 polarization detections were analyzed for their magnitude properties. Because the photometric values had only broad corrections on the magnitudes, as described in Section~\ref{polanal}, a post correction was implemented to obtain accurate magnitude values. This was carried out by zeropoint corrections using the 4453 2MASS matched stars, as described in Paper~I. The resultant photometry showed that the stars in GP182 region spanned magnitude ranges from $7\,\mathrm{mag}$ to $18\,\mathrm{mag}$. Errors in photometric magnitudes were below 1\% up to $13\,\mathrm{mag}$ stars and 10\% up to $15.5\,\mathrm{mag}$ stars. Figure~\ref{fig11} shows a plot of S/N against magnitude with the magnitude error in the right axis. The S/N values were obtained from the flux and flux error values of the stars. It is seen that the S/N achieved is greater than 10 for stars up to $15.5\,\mathrm{mag}$. This matched the desired signal-to-noise ratio goals described in Section~\ref{obgoals}.

The 4453 2MASS matched stars were compared with their magnitudes for estimating photometric accuracy. The difference in POLICAN and 2MASS magnitudes are plotted against their magnitude along with their corresponding magnitude errors in Figure~\ref{fig12}. The dispersion in magnitude differences were better than $0.05\,\mathrm{mag}$ up to $11\,\mathrm{mag}$ stars. For stars up to $13\,\mathrm{mag}$, the dispersion was around $0.1\,\mathrm{mag}$. For fainter stars, the dispersion increased to around $0.3\,\mathrm{mag}$, due to larger photometric uncertainties.

\begin{figure}[!ht]
\epsscale{1.1}
\plotone{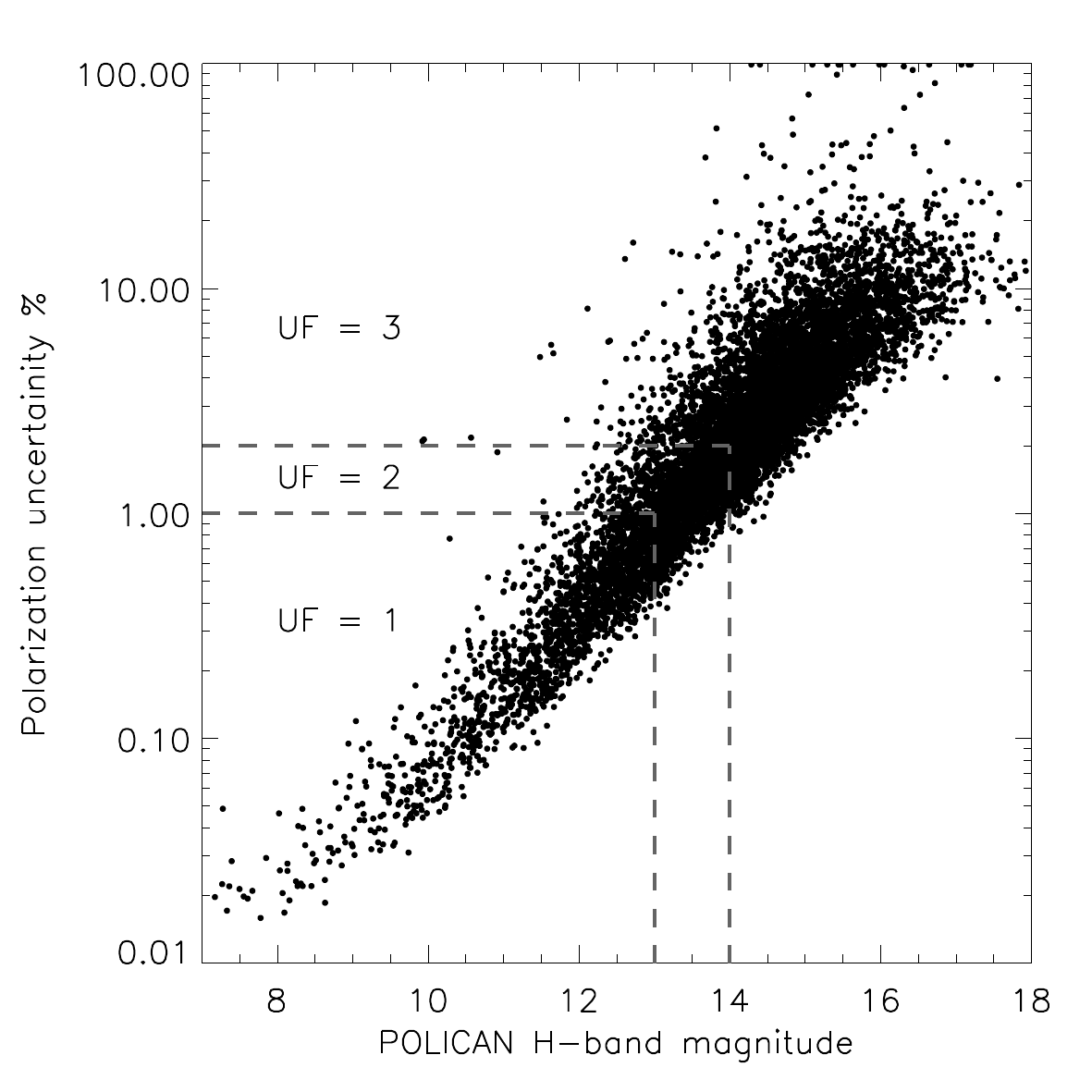} 
\caption{Plot of polarization uncertainty in log scale against POLICAN $H$-band magnitude for 9556 stars towards the GP182 region. The dashed line represents reliable polarization classification from UF~=~1 to 3 category. The high-quality polarization values are within UF~=~1 and have polarization uncertainties less than 1\% with magnitudes below $13\,\mathrm{mag}$.}
\label{fig13}
\end{figure}

\subsection{Polarimetric properties}

The polarimetric data available for all the 9556 stars provided a full range of polarization properties. Uncertainties in polarization ($\sigma_{P}$) values are useful to calculate the polarization S/N ($P_{S/N}$). Examining the polarization uncertainty against the magnitude showed POLICAN observations had polarization uncertainties of 1\% up to $13\,\mathrm{mag}$ and 2\% up to $14\,\mathrm{mag}$. This matched our polarization sensitivity goals as described in Section~\ref{obgoals}. Figure~\ref{fig13} shows the log plot of polarization uncertainty against POLICAN magnitude for stars from 7 to $18\,\mathrm{mag}$. 

\begin{deluxetable}{cc}
\tablecolumns{2}
\tablewidth{0pc}
\tablecaption{Polarimetric Usage Flags for POLICAN.\label{tbl-2}}
\tablehead{\colhead{Usage Flag} & {POLICAN data}}
\startdata
UF = 0 	 	&	$\sigma_{P}<1$ \begin{tiny}\&\end{tiny} mag$<$13\\	
			&	\begin{tiny}\&\end{tiny} $P_{S/N}>2.5$		   \\
UF = 1 	 	& 	$\sigma_{P}<1$ \begin{tiny}\&\end{tiny} mag$<$13\\
UF = 2  		&	$\sigma_{P}<2$ \begin{tiny}\&\end{tiny} mag$<$14\\
UF = 3 		&	$\sigma_{P}>2$ \begin{tiny}\&\end{tiny} mag$>$14\\				
\enddata
\end{deluxetable}

\begin{figure*}[!ht]
\epsscale{1.1}
\plotone{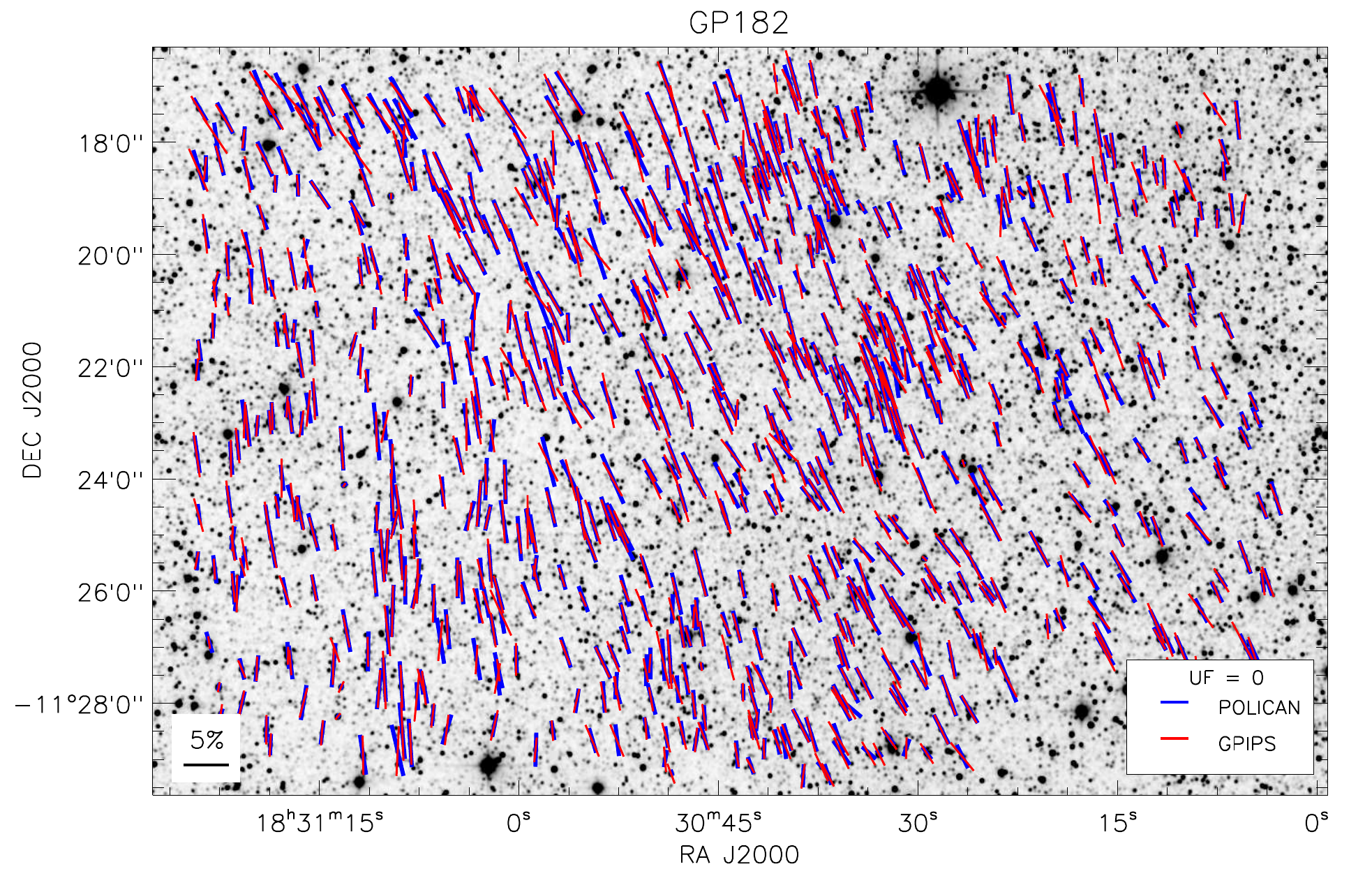} 
\caption{ The GP182 region is shown in the background image from the 2MASS $H$-band survey. The polarization values are plotted for common stars with UF~=~0 category matched in the GPIPS and POLICAN data. The blue vectors represent POLICAN observations and the red vectors represent GPIPS data. A reference vector in black at the bottom left indicates 5\% polarization.}
\label{fig14}
\end{figure*}

The reliability of polarization data can be determined from combination of $\sigma_{P}$, $P_{S/N}$ and magnitude. \citet{clemens12c} classified the stellar polarizations based on their reliability into usage flags (UF) to allow easy identifications. As magnetic field studies using POLICAN's starlight polarimetry have well-established observing goals and scheme (Section~\ref{obgoals} and \ref{obscheme}), it will be useful to classify POLICAN data. Based on the observed polarization properties, we formed a new set of usage flags for POLICAN data as follows: UF~=~1 represented the high-quality polarization values having $\sigma_{P}$ within 1\% and magnitude $<$ 13.0. Further, the stars with UF~=~1 having $P_{S/N}>2.5$ were categorized under UF~=~0 category. UF~=~0 represents the highest-quality of polarization values that can directly trace magnetic field directions with lowest dispersion in position angles. UF~=~2 sample is categorized for stars with $\sigma_{P}$ within 2\% and magnitude $<$ 14.0. They represent the moderately resolved magnetic field directions. The UF~=~2 sample can mostly be used to produce a mean magnetic field for a region with higher stellar density. The rest of stars belong to UF~=~3 category. They can be averaged with UF~=~1 and 2 stars to provide a very low resolution map of magnetic field. The average polarization values from UF~=~3 can also predict the mean polarization in large regions of the local ISM. Table~\ref{tbl-2} lists the complete classification of stars based on their usage flags for POLICAN. Figure~\ref{fig13} shows the usage flag classification on the plot of polarization uncertainty against magnitude.

\begin{figure*}[!ht]
\epsscale{1.0}
\plotone{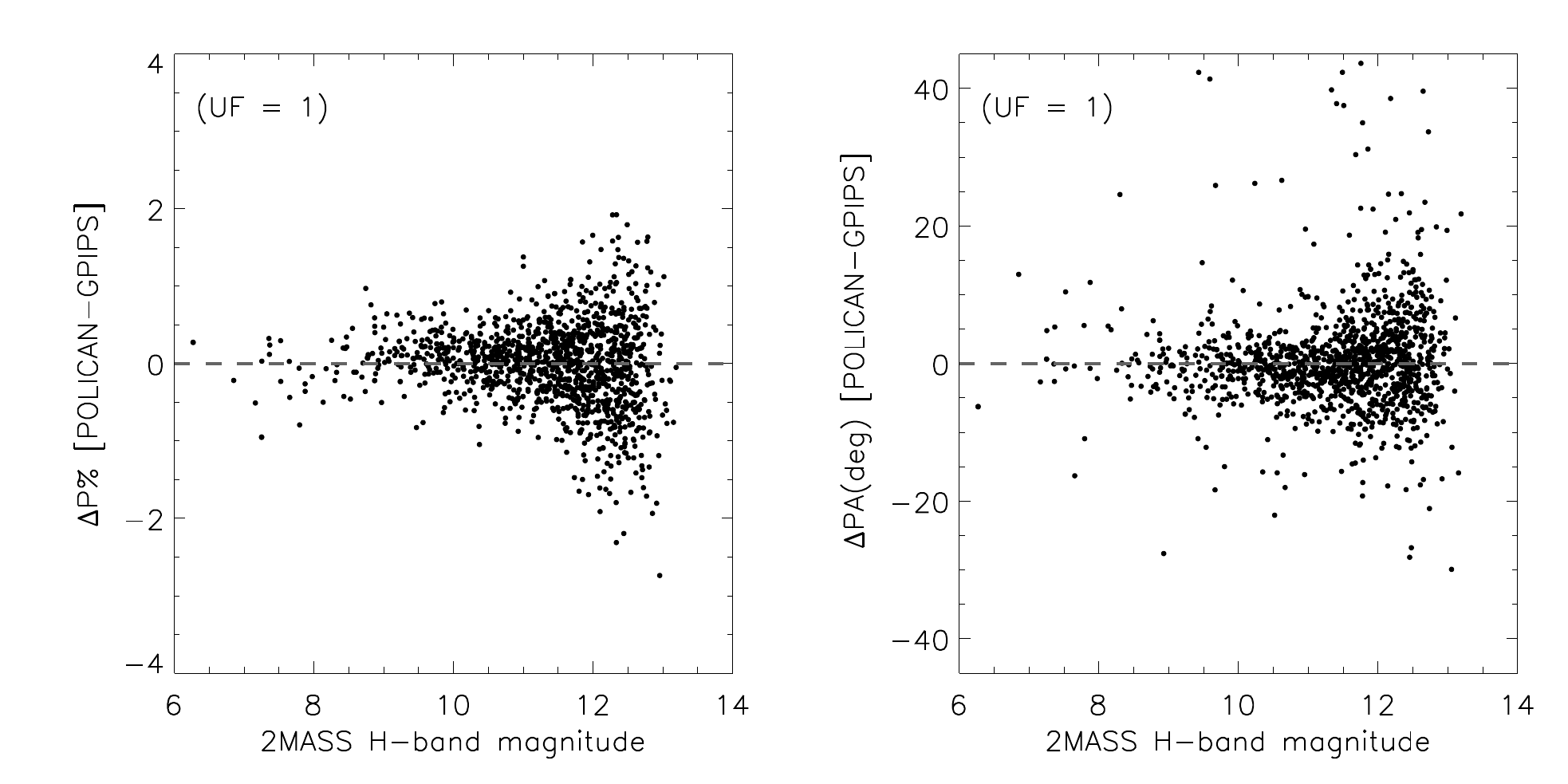} 
\caption{Plot of differences in GP182 polarization values between POLICAN and GPIPS data for 1298 stars with UF~=~1 category. The \textit{left panel} shows difference in degree of polarization against 2MASS $H$-band magnitude. The differences are within 1.5\% for majority of the stars. The \textit{right panel} shows differences in position angle against 2MASS $H$-band magnitude.}
\label{fig15}
\end{figure*}

\begin{figure*}[!ht]
\epsscale{1.0}
\plotone{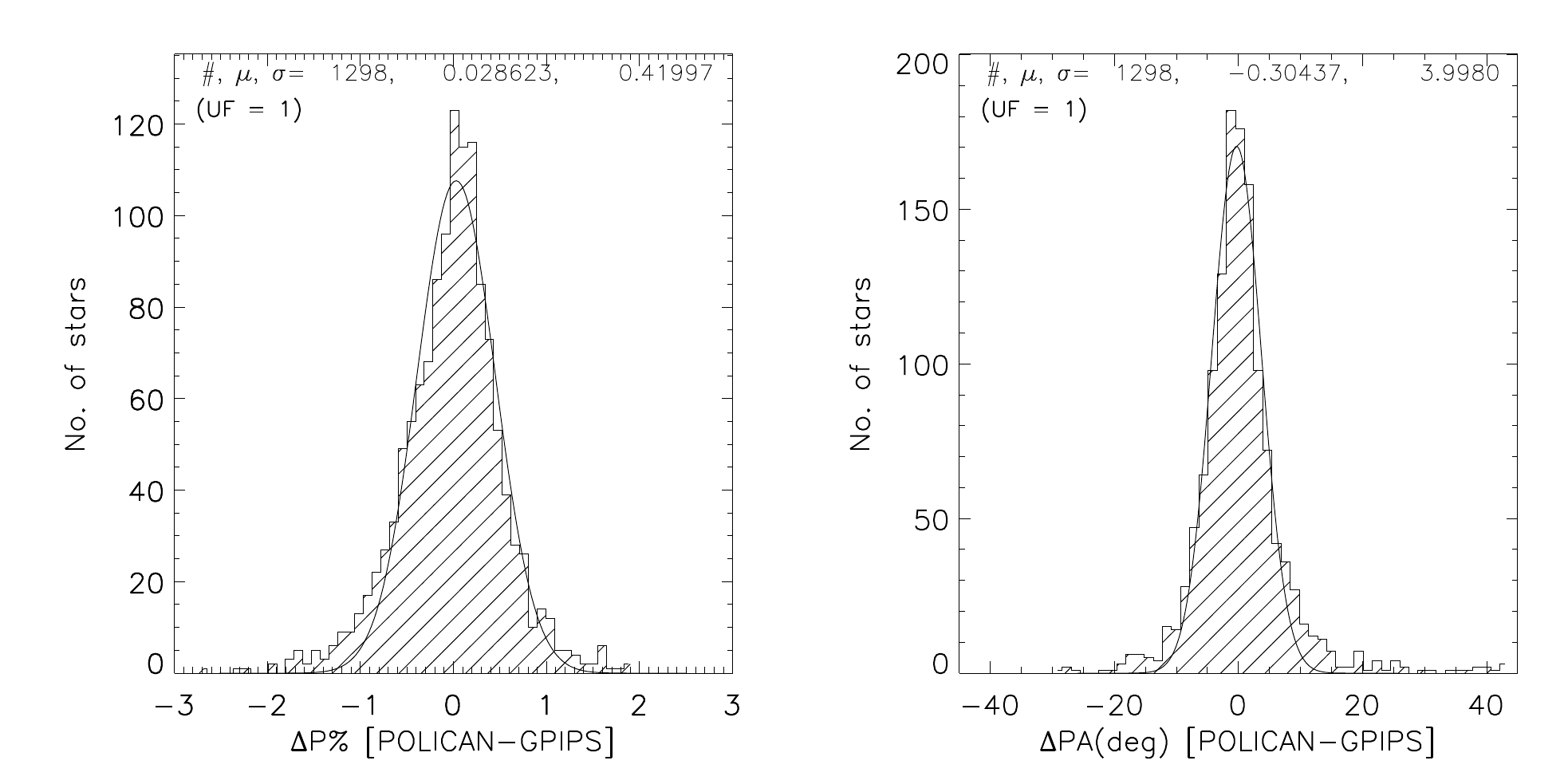} 
\caption{Histogram of differences in GP182 polarization values between POLICAN and GPIPS data for 1298 stars with UF~=~1 category. The histograms are fitted with a Gaussian to calculate the mean and standard deviation of the distribution. The polarization values have standard deviation better than 0.5\% with the position angle values below $5^{\circ}$.}
\label{fig16}
\end{figure*}

\section{Performance}\label{per}

After establishing POLICAN's GP182 stellar photometric and polarimetric properties, the results could be compared with the GPIPS data. The polarimetric data from GPIPS survey was derived from \textit{GPIPS data release 3.1}. This included the most recent version with the best compilation of up to date data. Individual stars from POLICAN and GPIPS data were matched to obtain common detections. A total of 1298 stars had common detection for UF~=~1 category, with also common detection in the 2MASS data. Out of these, there were 817 stars that matched UF~=~0 category. Figure~\ref{fig14} shows the background 2MASS image of GP182 region with polarization vectors for UF~=~0 stars. The polarization values of POLICAN are overlaid in blue vectors, with the GPIPS values overlaid in red vectors. Visually, the polarization vectors align with each other in their length and position angle. 

\begin{deluxetable*}{lccc}
\tablecolumns{3}
\tablewidth{0pc}
\tablecaption{Summary of POLICAN performance.\label{tbl-3}}
\tablehead{\colhead{Quantity} & \colhead{Value}    & \colhead{Comments}}
\startdata
Plate scale 					& 0.32 arcsec/pixel 			& On the detector \\
Field of view 				& $4\times4$ arcmin$^{2}$ 	& Cropped from $5.5\times5.5$ arcmin$^{2}$ \\
\hline
Photometric accuracy 		& $<0.1\,\mathrm{mag}$ 		& For stars brighter than $13\,\mathrm{mag}$\\	
Polarimetric accuracy 		& $<$ 0.5\% 					& For stars with UF = 1 \\	
Position angle accuracy 		& $<5^{\circ}$ 				& For stars with UF = 1 \\
Instrumental polarization 	& 0.51\% 		 			& $Q_{inst}=-0.50\%$ and $U_{inst}=0.12\%$\\	
Polarization uncertainties 	& 0.1\% to 10\%				& From $7\,\mathrm{mag}$ to $16\,\mathrm{mag}$\\
HWP zero-phase offset angle 	& $139^{\circ}$ 		 		& Correction angle\\
\enddata
\end{deluxetable*}

The values of UF~=~1 subset form the core information to determine POLICAN's polarimetric accuracy and performance. Since UF~=~1 are considered reliable and high-quality data, they need to equal the GPIPS data consistently. Comparison of both the data sets were carried out for all the 1298 matched stars. The differences between POLICAN and GPIPS stars were established by subtracting their individual polarization values. The difference in polarization percentage were below or around $0.5\%$ for stars brighter than $11\,\mathrm{mag}$. For fainter stars the difference reached up to $\sim1.5\%$. The position angle difference were below or around $5^{\circ}$ for stars brighter than $11\,\mathrm{mag}$. For fainter stars, the difference in position angle reached up to $\sim15^{\circ}$, with some exceeding it. Figure~\ref{fig15} shows the plot of POLICAN and GPIPS polarization and position angle differences against 2MASS $H$-band magnitude.

A histogram examination of the polarization differences gives the accuracy of POLICAN. In Figure~\ref{fig16} we show the histogram distribution for both polarization and position angle differences. A Gaussian is fitted for each distribution to determine the peak and standard deviation. The peak in the polarization and position angle differences are close to zero, indicating there is no offset in the calculated values. This means that the polarimetric efficiency, HWP zero-phase offset angle, instrumental polarization and polarization de-biasing, are well established for POLICAN. Because the data set represented high-quality UF~=~1 stars, the standard deviation of the Gaussian fit should give the polarimetric accuracy of POLICAN. From the fit, we see that the polarization accuracy is found to better than 0.5\% and the position angle accuracy is below $5^{\circ}$. The values of accuracy are minimum and within the expected levels of uncertainties with POLICAN data.

The established accuracies with POLICAN allow to obtain precise calculation of magnetic field strengths using \citet{chan53} method. Overall, POLICAN's UF~=~1 subset showed good agreement with archival data, adequately meeting the polarimetric goals for magnetic field studies.
 
\section{Summary}

We have described the important aspects in the operation, data processing, calibration, and performance of the newly developed polarimeter: POLICAN, at the $2.1\,\mathrm{m}$ OAGH telescope in M\'exico. POLICAN consists of a HWP and a polarizer that are housed in a mechanical assembly attached to CANICA and the telescope. The setup with plate scale of $0.32\,\mathrm{arcsec/pixel}$ and useful FOV of $4 \times 4\,\mathrm{arcmin^2}$ enables deep high-resolution medium field linear polarimetric imaging. The observation schemes are optimized to study polarization properties of both point sources and extended sources in the interstellar medium. POLICAN's large data sets of raw images are handled by the robust image reduction and analysis techniques implemented into custom pipelines in IRAF and IDL environment.

Polarimetric calibrations were carried out from observations of globular clusters and polarimetric standards. The analysis of 10,700 stars from 37 observations of globular cluster M5, determined the instrumental polarization to be 0.51\%. Observations of polarimetric standard HD38563C, determined the HWP zero-phase offset angle to be 139$^\circ$.  Pilot studies were carried out for both extended and point source regions to obtain POLICAN's observational results and performance. Scattered polarization was compared with SIRPOL data for Sh 2-106 object. Starlight polarimetry was compared with GPIPS data for GP182 field. Mapping a GP182 region of $20 \times 12\,\mathrm{arcmin^2}$ produced 9556 polarization detections reaching sensitivity many orders better than 2MASS survey. The polarimetric data were classified with usage flags to deem their reliability. A total of 1298 stars with reliable polarization under UF~=~1 category were compared with the GPIPS data. POLICAN achieved polarization accuracy better than 0.5\% and position angle errors below $5^\circ$ up to $13\,\mathrm{mag}$ stars in $H$-band. The entire performance of POLICAN is summarized in Table~\ref{tbl-3}. Based on background starlight polarimetry, POLICAN data can be used to trace the plane-of-sky magnetic field directions in the interstellar medium. Various observations on star-forming regions are being conducted towards the galactic plane to study their magnetic field properties. POLICAN features all characteristics of a sensitive NIR polarimeter capable of delivering reliable polarization data in the coming years.

\acknowledgements
We would like to thank the anonymous referee for useful comments on improving the article. We thank all the OAGH staff for their help in development and observations with the instrument. We are deeply indebted to the the valuable comments and feedback provided by Dan Clemens, Boston University, on improving the data processing techniques and calibration methods. We thank Eswaraiah Chakali, NTHU Taiwan, for helpful discussion on polarimetric analysis. This work has been carried out at Instituto Nacional de Astrof\'isica, \'Optica y Electr\'onica, M\'exico, with support from CONACyT under the project CB-2012-01 182841. D.R. with CVU 555629 acknowledges CONACyT for the grant 370405. SAOImage DS9 software is developed with the funding from the Chandra X-ray Science Center (CXC), the High Energy Astrophysics Science Archive Center (HEASARC) and JWST Mission office at Space Telescope Science Institute. This work makes use of data products from the Two Micron All Sky Survey, which is a joint project of the University of Massachusetts and the Infrared Processing and Analysis Center/California Institute of Technology, funded by the National Aeronautics and Space Administration and the National Science Foundation. This research used data from the Boston University (BU) Galactic Plane Infrared Polarization Survey (GPIPS), funded in part by NSF grants AST 06-07500, 09-07790, and 14-12269. GPIPS used the Mimir instrument, jointly developed at BU and Lowell Observatory and supported by NASA, NSF, and the W.M. Keck Foundation.

\facility{OAGH (CANICA, POLICAN)}.

\appendix

\section{Stokes parameters and Mueller matrices}\label{mlrmat}

The Stokes parameters define the polarization state of a non-coherent electromagnetic radiation. Originally described by G.~G.~Stokes in his classic paper \citet{stokes52}, the Stokes parameters were re-introduced to modern astronomy by \citet{chan47}, who denoted the polarization states by $I$, $Q$, $U$ and $V$.  The Stokes parameters can be combined to form a vector $S$ as
\begin{equation}
S = \begin{pmatrix}
I \\ Q \\ U \\ V
\end{pmatrix}
\end{equation}
where $I$ is the total intensity of the radiation; $Q$ is the intensity difference between horizontal and vertical linearly polarized components; $U$ is the intensity difference between linearly polarized components oriented at $\pm45^{\circ}$; and $V$ is the intensity of the circularly polarized radiation.

When electromagnetic radiation interacts with matter, it is likely to change its polarization state. The change in polarization state can be algebraically represented by matrix transformations of the input Stokes vector and the final measured Stokes vector. \citet{mueller48} described the matrix calculus for different states of polarization, each represented by its $4\times 4$ matrix transformation equation. The Mueller matrix formalism is always carried out by matrix multiplication in a particular order, which is from the final measured Stokes vector to the input vector. 

In the POLICAN setup, we use two polarizing components, which are the HWP and a polarizer. The matrix formalism for POLICAN with an input Stokes vector $S_{in}$ and the final measured Stokes vector $S_{out}$ can be represented as
\begin{equation}
S_{out} = M_{pol} * M_{HWP} * S_{in} 
\end{equation} 
where $M_{pol}$ is the Mueller matrix of the linear polarizer with fast axis oriented at $0^{\circ}$, and $M_{HWP}$ is the Mueller matrix of the HWP with arbitrary fast axis orientation $\theta$. These are represented by their respective Mueller matrices \citep{shur62} as

\begin{equation}
M_{pol} = \frac{1}{2}\begin{pmatrix}
1 & -1 & 0 & 0 \\
-1 & 1 & 0 & 0 \\
0 & 0 & 0 & 0 \\ 
0 & 0 & 0 & 0  \end{pmatrix}
\quad\mathrm{and}\quad
M_{HWP} = \begin{pmatrix}
1 & 0 & 0 & 0 \\
0 & cos(2\theta)^2-sin(2\theta)^2 & 2sin(2\theta)cos(2\theta) & 0 \\
0 & 2sin(2\theta)cos(2\theta) & cos(2\theta)^2-sin(2\theta)^2 & 0 \\   
0 & 0 & 0 & 0  \end{pmatrix}
\end{equation} 

Substituting the above matrices in equation~A2, we can write the final Stokes parameter $S_{out}$ as
\begin{equation}
S_{out} = \frac{1}{2}\begin{pmatrix}
I_{in} + Q_{in}(cos(2\theta)^2-sin(2\theta)^2) - U_{in}2sin(2\theta)cos(2\theta) \\
-I_{in} - Q_{in}(cos(2\theta)^2-sin(2\theta)^2) + U_{in}2sin(2\theta)cos(2\theta) \\  
0  \\ 
0   \end{pmatrix}
\end{equation} 

The final measured intensity $I_{out}$ of the Stokes vector $S_{out}$, depends on the HWP fast axis $\theta$. As noted in Section~\ref{opcntrl}, four HWP modulation angles are necessary for estimating the input linear Stokes parameters: $Q_{in}$ and $U_{in}$. Given that $\theta$ can have number of values between $0$ and 360$^{\circ}$, we can chose values such that $sine$ and $cosine$ functions in equation~A4 can cancel out to remain $Q_{in}$ and $U_{in}$. The first four angles of $\theta$ that fulfill the conditions are
\begin{equation}
\begin{matrix}
at & \theta = 0^{\circ} 	\\
& \theta = 22.5^{\circ}  \\
& \theta = 45^{\circ} 	\\
& \theta = 67.5^{\circ} 
\end{matrix} \quad\quad\quad
\begin{matrix}
sin(2\theta) = 0 \quad\mathrm{and}\quad cos(2\theta) = 1 			\\
sin(2\theta) = 1/\sqrt{2} \quad\mathrm{and}\quad cos(2\theta) = 1/\sqrt{2}   \\
sin(2\theta) = 1 \quad\mathrm{and}\quad cos(2\theta) = 0  			 \\
sin(2\theta) = 1/\sqrt{2} \quad\mathrm{and}\quad cos(2\theta) = -1/\sqrt{2} 
\end{matrix}
\end{equation}

Substituting these values in equation~A4 and combining the Stokes vector $S_{out}$ as $I_{out}$, we get the output intensity for each HWP angle as
\begin{equation}
\begin{matrix}
I_{0} = \frac{1}{2}[\pm I_{in}\mp Q_{in}] \\
I_{22.5} = \frac{1}{2}[\pm I_{in}\mp U_{in}] \\
I_{45} = \frac{1}{2}[\pm I_{in}\pm Q_{in}] \\
I_{67.5} = \frac{1}{2}[\pm I_{in}\pm U_{in}] \\
\end{matrix}
\end{equation}

Now, we can re-arrange equation~A6 to establish the input Stokes parameters as

\begin{equation}
\begin{matrix}
I_{in} = (I_{0} + I_{45} +I_{22.5} + I_{67.5})/2 \\
Q_{in} = I_{0} - I_{45} \\
U_{in} = I_{22.5} - I_{67.5}
\end{matrix}
\end{equation}

\section{Polarimetric Error analysis}\label{polerr}

Polarimetric analysis of point sources (mainly stars) are obtained by measuring the fluxes (integrated counts) on the stars in images corresponding to each of the orthogonal polarized components of linear Stokes parameters. The flux measurement in POLICAN is performed through synthetic aperture photometry on brightness profiles of the stars in the observed images based on the use of DAOPHOT package in IDL (see Section~\ref{polanal}).

The \textit{phot/aper} function of the DAOPHOT package applied to the image with stars, measures the flux of a source in values of analog-to-digital units (ADU) as follows \citep{stet87}:
\begin{equation}
I_{s} = I_{tot} - (area*skymod)
\end{equation}
where $I_{s}$ is the total flux measured of the source within an aperture, $I_{tot}$ is the total flux measured within an aperture, $area$ is the total area of pixels in the aperture, and $skymod$ is the sky/background modal value per pixel measured from all the pixel values within a sky annulus (In IDL this is obtained by \textit{mmm}).

The error in flux measurement $\sigma_{s}$ in ADU is given as follows:
\begin{equation}
\sigma_{s} = \sqrt{(\frac{I_{s}}{gN_{i}}) + (area*skyvar) + (\frac{area^2*skyvar}{nsky})}
\end{equation}
where $skyvar$ is the variance in sky measurement per pixel for the final image, $g$ is the gain in electrons/ADU, $N_{i}$ is the number of images used for constructing the final image, and $nsky$ is the number of pixels used in the sky annulus during photometry. 

Based on the equation of Stokes parameters as described in Section~\ref{polanal} and Appendix~\ref{mlrmat}, the error in Stokes parameters can be given by standard error propagation as follows:

\begin{equation}
\sigma_{I} = \frac{\sqrt{\sigma_{0}^2 + \sigma_{22.5}^2 + \sigma_{45}^2 + \sigma_{67.5}^2}}{2}
\end{equation}
\begin{equation}
\sigma_{Q} = \sqrt{\frac{\sigma_{0}^2 + \sigma_{45}^2}{I^2} + (\frac{Q}{I}\sigma_{I})^2}
\end{equation}
\begin{equation}
\sigma_{U} = \sqrt{\frac{\sigma_{22.5}^2 + \sigma_{67.5}^2}{I^2} + (\frac{U}{I}\sigma_{I})^2}
\end{equation}

where $\sigma_{0}, \sigma_{22.5}, \sigma_{45}, \sigma_{67.5}$ are flux errors for the fluxes $I_{0}, I_{22.5}, I_{45}, I_{67.5}$ measured in each HWP angle.

The Stokes parameters are next scaled by polarization efficiency $\eta$ and rotated by the HWP zero-phase offset angle, $\theta$ as shown in equation~\ref{eqn5}. The corresponding Stokes errors in equatorial system are
\begin{equation}
\sigma_{Qeq} = \sqrt{(\frac{cos2\theta}{\eta}\sigma_{Q})^2 + (\frac{sin2\theta}{\eta}\sigma_{U})^2}
\end{equation}
\begin{equation}
\sigma_{Ueq} = \sqrt{(\frac{cos2\theta}{\eta}\sigma_{U})^2 + (\frac{sin2\theta}{\eta}\sigma_{Q})^2}
\end{equation}

Next, the Stokes values are corrected for instrumental polarization as in equation~\ref{eqn7} and are represented with their errors as
\begin{equation}
\sigma_{Qc} = \sqrt{\sigma_{Qeq}^2 + \sigma_{Qinst}^2}
\end{equation}
\begin{equation}
\sigma_{Uc} = \sqrt{\sigma_{Ueq}^2 + \sigma_{Uinst}^2}
\end{equation}
where $\sigma_{Qinst}$ and $\sigma_{Uinst}$ are error in instrumental Stokes values calculated from globular cluster observations (see Section~\ref{pcalgc}).

The final Stokes values are combined to give the equatorial degree of polarization $P_{eq}$ and the position angle $P.A$ as in equation~\ref{eqn8}~and~\ref{eqn9}. The error in polarization is calculated as
\begin{equation}
\sigma_{P} = \frac{100^2}{P_{eq}}\sqrt{(Q_{c}^2*\sigma_{Qc}^2) + (U_{c}^2*\sigma_{Uc}^2)}
\end{equation}

After obtaining the de-biased polarization value $P$ as in equation~\ref{eqn10}, the $P.A.$ uncertainty ($\sigma_{P.A}$) is computed as
\begin{equation}
\sigma_{P.A} = 28.65(\frac{\sigma_{P}}{P})
\end{equation}


\begin{thebibliography}{}

\bibitem[Carrasco et al.(2017)]{car17} Carrasco, L., Hern\'{a}ndez Utrera, O., V\'{a}zquez, S., et al.\ 2017, \rmxaa, 53, 497

\bibitem[Chandrashekar(1947)]{chan47} Chandrasekhar, S.\ 1947, \apj, 105, 424

\bibitem[Chandrashekar \& Fermi(1953)]{chan53} Chandrasekhar, S., \& Fermi, E.\ 1953, \apj, 118, 113 

\bibitem[Chapman et al.(2011)]{chapman11} Chapman, N. L., Goldsmith, P. F., Pineda, J. L., Clemens, D. P.\ 2011, \apj, 741, 21 

\bibitem[Clarke (2010)]{clarke10} Clarke, David\ 2010, Stellar Polarimetry, Wiley-VCH publications

\bibitem[Clemens et al.(2007)]{clemens07} Clemens, D. P., Sarcia, D., Grabau, A., et al.\ 2007, \pasp, 119, 1385 

\bibitem[Clemens et al.(2012a)]{clemens12a} Clemens, D. P., Pinnick, A., Pavel, M. D., \& Taylor, B. W.\ 2012, \apjs, 200, 19 

\bibitem[Clemens et al.(2012b)]{clemens12b} Clemens, D. P., Pinnick, A., Pavel, M. D.\ 2012, \apjs, 200, 20 

\bibitem[Clemens et al.(2012c)]{clemens12c} Clemens, D. P., Pavel, M. D., \& Cashman, L. R.\ 2012, \apjs, 200, 21

\bibitem[Clemens et al.(2013)]{clemens13} Clemens, D. P., Pavel, M. D., \& Cashman, L. R.\ 2013, \apjs, 145, 74 

\bibitem[Davis \& Greenstien(1951)]{davis51} Davis, L., Jr., \& Greenstein, J. L.\ 1951, \apj, 114, 206

\bibitem[Devaraj et al.(2015)]{dev15} Devaraj, R., Luna, A., Carrasco, L., \& Mayya, Y. D.\ 2015, IAU Symposium 305, 10, 175

\bibitem[Devaraj et al.(2017a)]{dev17a} Devaraj, R., Luna, A., Carrasco, L., \& Mayya, Y. D.\ 2017, \rmxaa(Conf. Series), 49, 58

\bibitem[Devaraj et al.(2017)]{dev17} Devaraj, R., Mayya, Y. D., Carrasco, L., et al.\ 2017, \pasp, Accepted (Paper~I)

\bibitem[Hall(1949)]{hall49} Hall, J. S.\ 1949, Science, 109, 166

\bibitem[Hashimoto et al.(2008)]{hashimoto08} Hashimoto, J., Tamura, M., Kandori, R., et al.\ 2008, \apjl, 677, L39

\bibitem[Hiltner(1949)]{hiltner49} Hiltner, W. A.\ 1949, \apj, 109, 471

\bibitem[Howell(1989)]{howell89} Howell, Steve B.\ 1989, \pasp, 101, 616

\bibitem[Hough et al.(2006)]{hough06} Hough, J. H., Lucas, P. W., Bailey, J. A., et al.\ 2006, \pasp, 118, 1302

\bibitem[Jones(1989)]{jones89} Jones, Terry Jay\ 1989, \apj, 346, 728 

\bibitem[Jones(1997)]{jones97} Jones, Terry Jay\ 1997, \aj, 114, 1393

\bibitem[Kandori et al.(2006)]{kandori06} Kandori, R., Tamura, M., Kusakabe, N., et al.\ 2006, \procspie, 6269, 51

\bibitem[King et al. (2014)]{king14} King, O. G., Blinov, D., Ramaprakash, A. N., et al.\ 2014, \mnras, 442, 1706

\bibitem[Landsman(1993)]{land93} Landsman, W. B.\ 1993, ASP Conf. Series, 52, 246

\bibitem[Lang et al.(2010)]{lang10} Lang, D., Hogg, D. W., Mierle, K., Blanton, M., \& Roweis, S.\ 2010, \aj, 139, 1782

\bibitem[Lazarain \& Hoang(2007)]{lazarain07} Lazarian, A., \& Hoang, T.\ 2007, \mnras, 378, 910

\bibitem[Magalh\~{a}es et al. (1996)]{mag96} Magalh\~{a}es, A. M., Rodrigues, C. V., Margoniner, V. E., et al.\ 1996, ASP Conf. Series, 97, 118

\bibitem[Masiero et al.(2007)]{masiero07} Masiero, J., Hodapp, K., Harrington, D., \& Haosheng Lin\ 2006, \pasp, 119, 1126

\bibitem[Mathewson \& Ford(1970)]{mat70} Mathewson, D. S. \& Ford, V. L. 1970, \memras, 74, 139 

\bibitem[Montgomery \& Clemens(2014)]{mont14} Montgomery, J. D., \& Clemens, D. P.\ 2014, \apj, 786, 41

\bibitem[Mueller(1948)]{mueller48} Mueller, H.,\ 1948, The foundations of Optics, \josa, 38, 661

\bibitem[Nishiyama et al.(2009)]{nishi09} Nishiyama, S., Tamura, M., Hatano, H., Kanai, S., et al.\ 2009, \apjs, 690, 1648

\bibitem[Packham et al. (2012)]{pack12} Packham, C., Jones, T. J., Warner, C., et al.\ 2012, \procspie, 8446, 3R

\bibitem[Perrin et al.(2015)]{perrin15} Perrin, M. D., Duchene, G. Millar-Blanchaer, M., et al.\ 2015, \apj, 799, 182

\bibitem[Piirola et al.(2014)]{pir14} Piirola V., Berdyugin A., Berdyugina S., 2014, \procspie, 9147, 8I

\bibitem[Ramaprakash et al.(1998)]{ram98} Ramaprakash A. N., Gupta R., Sen A. K., Tandon S. N., 1998, \aaps, 128, 369

\bibitem[Roelfsema et al.(2010)]{rol10} Roelfsema, R., Martin, H., Schmid, H. M., et al.\ 2010, \procspie, 7735, 4B

\bibitem[Skrutskie et al.(2006)]{skrut06} Skrutskie, M. F., Cutri, R. M., Stiening, R., et al.\ 2006, \aj, 131, 1163

\bibitem[Saito et al.(2009)]{saito09} Saito, H., Tamura, M., Kandori, R., Kusakabe, N., et al.\ 2009, \aj, 137, 3149

\bibitem[Shurcliff(1962)]{shur62} Shurcliff, W. A.\ 1962, Polarized Light: Production and Use, Harvard University Press

\bibitem[Stetson(1987)]{stet87} Stetson, P. B.\ 1987, \pasp, 99, 191

\bibitem[Stetson(1990)]{stet90} Stetson, P. B.\ 1990, \pasp, 102, 932

\bibitem[Stokes(1852)]{stokes52} Stokes, G. G.\ 1852, Transactions of Cambridge Philosophical Society, 399-416 

\bibitem[Tamura et al.(2006)]{tamura06} Tamura, M., Kandori, R., Kusakabe, N., et al.\ 2006, \apjl, 649, L29

\bibitem[Vacca et al.(2004)]{vacca04} Vacca, W. D., Cushings, M. C., \& Rayner, J. T.\ 2004, \pasp, 116, 352

\bibitem[V\'{a}zquez-Rodr\'{i}guez(2012)]{vazrod12} V\'{a}zquez-Rodr\'{i}guez, M.\ 2012, INAOE, \url{http://inaoe.repositorioinstitucional.mx/jspui/handle/1009/779}

\bibitem[Wardle \& Kronberg(1974)]{wardle74} Wardle, J. F. C., \& Kronberg, P. P. 1974, \apj, 194, 249

\bibitem[Whittet et al.(1992)]{whittet92} Whittet, D. C. B., Martin, P. G., Hough, J. H., et al. 1992, \apj, 386, 562

\end{thebibliography}
\end{document}